\documentclass{article}

\usepackage{amsthm,amsmath,amssymb,amsfonts}
\usepackage{epsfig}
\usepackage{psfrag,eepic,mathtools}
\usepackage{url,hyperref}
\usepackage{ytableau}
\usepackage{tikz}

\makeatletter \@addtoreset{equation}{section}

\makeatother

\newcommand{\seteq}{\mathbin{:=}}
\newcommand{\simto}{\xrightarrow{\,\sim\,}}
\newcommand{\C}{{\mathbb C}}
\newcommand{\Z}{{\mathbb Z}}
\newcommand{\eps}{\epsilon}
\newcommand{\la}{\langle}
\newcommand{\ra}{\rangle}

\newcommand{\Vir}{\mathrm{Vir}}
\newcommand{\NSR}{\textsf{NSR}}
\newcommand{\Asl}{\widehat{\mathfrak{sl}}}
\newcommand{\rmF} {\mathrm{F}}
\newcommand{\calM}{\mathcal{M}}
\newcommand{\calA}{\mathcal{A}}
\newcommand{\calF}{\mathcal{F}}
\newcommand{\calL}{\mathcal{L}}

\newcommand{\calE}{\mathcal{E}}
\newcommand{\calV}{\mathcal{V}}
\newcommand{\calH}{\mathcal{H}}
\newcommand{\sfV}{\textsf{V}}

\makeatletter
\def\Maketitle{{\def\newpage{}\maketitle}}
\def\Appendix{\appendix
  \def\@seccntformat##1{Appendix~\csname the##1\endcsname.~~}}
\makeatother
\def\secpart{\refstepcounter{subsubsection}
\par\smallskip\par\noindent\textbf{\thesubsubsection.~}}
\def\sectpart{\refstepcounter{subsection}
\par\medskip\par\noindent\textbf{\thesubsection.~}}

\newtheorem{Proposition}{Proposition}[section]
\newtheorem{lemma}{Lemma}[section]

\newtheorem{Conj}{Conjecture}[section]

\begin{document}

\large

\title{\textbf{Bases in coset conformal field theory from AGT correspondence and Macdonald polynomials at the roots of unity}}
\author{A.~A.~Belavin$^{1,2,3}$, M.~A.~Bershtein$^{1,2,3,4}$ and G.~M.~Tarnopolsky$^{1,3,5}$\vspace*{10pt}\\[\medskipamount]
$^1$~\parbox[t]{0.88\textwidth}{\normalsize\it\raggedright
Landau Institute for Theoretical Physics,
Chernogolovka, Russia}\vspace*{4pt}\\[\medskipamount]
$^2$~\parbox[t]{0.88\textwidth}{\normalsize\it\raggedright Institute for Information Transmission Problems,  Moscow, Russia}\vspace*{4pt}\\[\medskipamount]
$^3$~\parbox[t]{0.88\textwidth}{\normalsize\it\raggedright Moscow Institute of Physics and Technology, Dolgoprudny, Russia}\vspace*{4pt}\\[\medskipamount]
$^4$~\parbox[t]{0.88\textwidth}{\normalsize\it\raggedright
Independent University of Moscow, Moscow,
Russia}\vspace*{4pt}\\[\medskipamount]
$^5$~\parbox[t]{0.88\textwidth}{\normalsize\it\raggedright
Department of Physics, Princeton University, NJ, USA}}
\date{}
\rightline{\texttt{\today}}
\Maketitle

\begin{abstract}\vspace*{2pt}
We continue our study of the AGT correspondence
between instanton counting on $\mathbb{C}^2/\Z_p$ and Conformal field theories with the symmetry algebra $\calA(r,p)$. In the cases $r=1, p=2$ and $r=2, p=2$ this algebra specialized to:  $\calA(1,2)=\calH \oplus \Asl(2)_1$ and $\calA(2,2)=\calH \oplus \Asl(2)_2 \oplus \NSR$.

As the main tool we use a new construction of the algebra $\calA(r,2)$ as the limit of the toroidal $\mathfrak{gl}(1)$ algebra for $q,t$ tend to $-1$.  We claim that the basis of the representation of the algebra $\calA(r,2)$ (or equivalently, of the space of the local fields of the corresponding CFT) can be expressed through Macdonald polynomials with the parameters $q,t$ go to $-1$. The  vertex operator which naturally arises in this construction has factorized matrix elements in this basis. We also argue that the singular vectors of the $\mathcal{N}=1$ Super Virasoro algebra can be realized in terms of Macdonald polynomials for a rectangular Young diagram and parameters $q,t$ tend to $-1$.
\end{abstract}

\large
\section{Introduction}

\sectpart This paper is a sequel to \cite{BBFLT:2011}. We study some questions concerning conformal field theory (CFT) using (mainly for motivation) the AGT relation (\cite{AGT:2009}). Namely, the AGT relation between Conformal field theory with the symmetry algebra $\calA$ and Instanton moduli space $\calM$ suggests the following structure:
\begin{itemize}
\item Every torus fixed point $p\in\mathcal{M}$ corresponds to basic vector $v_p \in \pi_{\mathcal{A}}$ (highest weight representation of the algebra $\mathcal{A}$).
\item The basis $\{v_p\}$ is orthogonal under the natural product and the norm of the vector $v_p$ equals the determinant of the torus action in the tangent space of $p$.
\item Matrix elements of geometrically defined vertex operators have completely factorized form. The last expressions are also denoted by $Z_{\textsf{bif}}$ (contribution of the bifundamental multiplet).
\item There exists a commutative algebra (Integrals of Motion) which is diagonalized in the basis $\{v_p\}$.
\end{itemize}
For more details and motivation of this structure see \cite{Alday:2010vg}, \cite{Alba:2010qc} and \cite{BBFLT:2011}. In this paper we construct this basis for certain conformal filed theories.

\sectpart It was suggested in the papers \cite{Belavin_Feigin:2011},\cite{Nishioka:2011jk} that CFT with the symmetry
\begin{align}\label{eq:A(r,p)}
\mathcal{A}(r,p)\overset{\text{def}}{=}\mathcal{H}
\times \widehat{\mathfrak{sl}}(p)_r\times \frac{\widehat{\mathfrak{sl}}(r)_p \times \widehat{\mathfrak{sl}}(r)_{n-p}}{\widehat{\mathfrak{sl}}(r)_n}
\end{align} corresponds to the instanton counting on $\C^2/\Z_p$, where $\Z_p$ acts by formula $(z_1,z_2) \mapsto (\omega z_1,\omega^{-1}z_2)$, $\omega=\exp(\frac{2\pi i}p)$. The quotient $\C^2/\Z_p$ is singular, and there are several possibilities to define instanton moduli space. One of them can be performed as follows. Denote by $\mathcal{M}(r,N)$ the smooth compactified moduli space of $U(r)$ instantons on $\mathbb{C}^2$ with the topological number $N$. The set $\mathcal{M}(r,N)^{\mathbb{Z}_{p}}$ of $\mathbb{Z}_p$-invariant  points on $\mathcal{M}(r,N)$ is a smooth compactification of the space of instantons on $\mathbb{C}^2/\mathbb{Z}_p$.

Torus fixed points for this compactification are labeled by the $r$-tuples of Young diagrams $\vec{\lambda}^\sigma=(\lambda_1^{\sigma_1},\dots,\lambda_r^{\sigma_r})$ colored in $p$ colors, where $\sigma_j=0,\dots,p-1$ is a color of the angle of $\lambda_j$. Using this compactification the expression for the 4-point conformal block for the algebras $\calA(2,2)$ and $\calA(2,4)$ was proposed in \cite{BBB:2011tb} and \cite{Alfimov:2011ju}. This result suggests the existence of the basis in the representation $\calA(r,p)$ which is labeled by $\vec{\lambda}^\sigma$  such that matrix elements of the vertex operator have the form:
\begin{align}
\langle J_{\vec{\lambda}^{\tilde{\sigma}}}|\Phi_\alpha|J_{\vec{\mu}^{\sigma}}   \rangle=
\prod_{S(\lambda_i,\mu_j)}(-b l_{\lambda}(s)+b^{-1}a_{\mu}(s)+b^{-1}-\alpha-P_i+\tilde{P_j}) \, \cdot \, \notag \\ \cdot \prod_{S(\mu_j,\lambda_i)}(bl_{\mu}(t)+b-b^{-1}a_{\lambda}(t)+\alpha-P_i+\tilde{P}_j),
\label{eq:matrix:color}
\end{align}
where
\begin{equation*}
s\in S(\lambda^{\tilde{\sigma}}, \mu^{\sigma})  \Longleftrightarrow s\in \lambda,\;\text{and}\;
l_\lambda(s)+a_\mu(s)+1+\sigma-\tilde{\sigma}\equiv 0 \bmod p,
\end{equation*}
$a_\lambda$ and $l_\mu$ are the arm and leg length correspondingly, $P_i$ and $\tilde{P}_j$ highest weights of the representations of $\calA(r,p)$, $b$ is related to the  central charge of the algebra $\calA(r,p)$, $\alpha$ is a parameter of the vertex operator.

In this paper we construct this basis for the algebras $\calA(1,2)$ and $\calA(2,2)$. It follows from the coset formula \eqref{eq:A(r,p)} that
\begin{align*}
\calA(1,2)=\calH\oplus \Asl(2)_1,\qquad \calA(2,2)=\calH\oplus \Asl(2)_2\oplus \NSR,
\end{align*}
where $\calH$ is a Heisenberg algebra, $\Asl(2)_k$ is an affine algebra $\Asl(2)$ on the level $k$ and \textsf{NSR} is the Neveu--Schwarz--Ramond algebra (the $\mathcal{N}=1$ super--Virasoro algebra).

\sectpart Denote by $\calE_1(q,t)$ the quantum toroidal $\mathfrak{gl}(1)$ algebra \cite{Feigin_Miwa_2011} depending on two parameters $q,t$ (another names are Ding-Iohara algebra, see \cite{AFHKSY:2011}, or elliptic Hall algebra, see \cite{ShiffVass}). We conjecture that in the limit $q, t \rightarrow -1$ of level $r$ representations of $\calE_1(q,t)$ we will have the representations of the conformal algebra $\calA(r,2)$. For example one can take  the basis in the level $r$ representations of $\calE_1(q,t)$ introduced in \cite{AFHKSY:2011} and obtain in the limit the basis in the representations of the algebra $\calA(r,2)$. For the $r=1$ case one can explicitly see the algebra $\calA(1,2)$ in the limit of $\calE_1(q,t)$. For $r=2$ case we didn't prove the conjecture but support it by checks of three consequences: coincidence of characters, formulas for singular vectors and matrix elements of vertex operators.

One remark is in order. Denote by $\calE_2(q,t)$ the quantum toroidal $\mathfrak{gl}(2)$ algebra. Then it is expected that in the conformal limit $q,t \rightarrow 1$ of level $r$ representations of $\calE_2(q,t)$ we will also get conformal algebra $\calA(r,2)$. Thus these algebras have different constructions as the limit of the deformed algebras. These conjectures have obvious generalizations for $p>2$.

\sectpart Let us describe the results of our paper. For the $r=1$ the basis in the representation of $\calE_1(q,t)$ identified with the Macdonald polynomials $J_\lambda(q,t)$. Thus our basis identifies with the limit $q,t \rightarrow -1$ of the Macdonald polynomials. Such  polynomials were introduced by Uglov in \cite{Uglov 1,Uglov 2}, where he proved that these polynomials describe the eigenvectors of the spin generalization of the Calogero-Sutherland Model \cite{BGHP:1993}. We call these polynomials \emph{Uglov polynomials} as in \cite{Kuramoto Kato}. It is worth to note that our idea of the limit $q,t \rightarrow -1$ of the algebra $\calE_1(q,t)$ was based on the Uglov work. In addition to the Uglov results we see that the matrix elements of natural vertex operators \eqref{eqPhi_0_in_h},\eqref{eqW_h_n} has factorized form \eqref{eq:matrix:color}.

As was mentioned before in the limit of $\calE_1$ we can see the algebra $\calA(1,2)=\calH\oplus\Asl(2)_1$. The last algebra comes in the principal (Lepowsky-Wilson) realization. Note that in the limit of toroidal algebra $\calE_2(q,t)$ the algebra $\calA(1,2)=\calH\oplus\Asl(2)_1$ comes in homogenous (Frenkel-Kac) realization.

For the $r=2$ case we can take the limit of the basis from \cite{AFHKSY:2011} in a similar way. The obtained basis $J_{\vec{\lambda}^\sigma}$ is labeled by pair of Young diagrams $\lambda_1^{\sigma_1},\lambda_2^{\sigma_2}$ colored in two colors and (conjecturally) has the following properties:
\begin{itemize}
  \item If $\lambda_2=\varnothing$ then $J_{\lambda_1^{\sigma_1},\varnothing}$ is given in terms of Uglov polynomials after bosonization of algebra $\calA(2,2)=\calH\oplus \Asl(2)_2\oplus \NSR$ (see Conjecture \ref{Conj:J:varnohing}). Analogous facts for the algebra $\calA(2,1)=\calH\oplus \Vir$ were proved in~\cite{Alba:2010qc}.

In the same way if $\lambda_1=\varnothing$ then $J_{\varnothing,\lambda_2^{\sigma_2}}$ is given in terms of Uglov polynomials but after \emph{another} bosonization of $\calA(2,2)$. In the special case  $c_{\NSR}=\frac32$ (in CFT notation $Q=b+\frac1b=0$ or in notation of $\Omega$-deformation $\eps_1+\eps_2=0$) $J_{\vec{\lambda}^\sigma}$ is a product of two Uglov polynomials corresponding to diagrams $\lambda_1^{\sigma_1}$ and $\lambda_2^{\sigma_2}$.

  \item If $\lambda_2=\varnothing$, $\lambda_1$ is a rectangle $m\times n$, $m\equiv n (\textrm{mod}\, 2) $ and a highest weight of the representation $\calA(2,2)$ take special value $P_{m,n}$ then the vectors $J_{\lambda_1^{\sigma_1},\varnothing}$ become the singular vectors for the $\NSR$ algebra. This singular vectors coincide with the Uglov polynomials after some nontrivial change of variables (see Conjecture \ref{Conj:Sing}). This fact was checked up to the level $9/2$. This result is a analogue of the description of the singular vectors for the Virasoro algebra in terms of the Jack polynomials \cite{MimachiYamada1995,AMOS:1995} \footnote{see also \cite{Dobrev:1992} for the earlier results in this direction}.

  \item The matrix elements of the vertex operator $\Phi=\calV_\alpha\cdot \Phi_{\alpha}^{\scriptscriptstyle{\textsf{NS}}}$ have explicit factorized form \eqref{eq:matrix:color}, where $\calV_\alpha$ is ``rotated'' Heisenberg vertex \eqref{eqPhi_0_in_h} and $\Phi_{\alpha}^{\scriptscriptstyle{\textsf{NS}}}$ is the Neveu--Schwarz primary field.  This fact was checked up to level 2 (see Conjecture \ref{Conj:r2ME}). The expression \eqref{eq:matrix:color} also coincides with the limit of the matrix elements for the vertex of the algebra $\calE_1(q,t)$. Thus we expect that the limit of the last vertex is the operator $\Phi=\calV_\alpha \cdot \Phi_{\alpha}^{\scriptscriptstyle{\textsf{NS}}}$.
\end{itemize}

\newpage
\tableofcontents

\section{$r=1$ case} \label{seqp=2}

\subsection{Algebra, homogeneous realization and characters}
\secpart
In this case we study the conformal field theory with the symmetry algebra $\calA(1,2)=\calH\oplus \Asl(2)_1$. We denote the generators of the Heisenberg algebra $\calH$ by $w_n$, where $n \in \Z$ and relations have the form:
\begin{align*}
[w_n,w_m]=2n\delta_{n+m}.
\end{align*}

The Lie algebra $\Asl(2)=\mathfrak{sl}(2)\otimes \mathbb{C}[t,t^{-1}]\oplus \mathbb{C}K$ has generators $e_n=e\otimes t^n$, $f_n=f\otimes t^n$, $h_n=h\otimes t^n$ and the central generator $K$ with the relations:
\begin{align*}
[e_n,e_m]=[f_n,f_m]=0,\quad [e_n,f_m]=h_{n+m}+n\delta_{n+m}K, \\
[h_n,e_m]=2e_{n+m},\quad [h_n,f_m]=-2f_{n+m},\quad [h_n,h_m]=2n\delta_{n+m}K. \notag
\end{align*}

Denote by $\mathcal{L}_{h,k}$ the integrable representation of the $\widehat{\mathfrak{sl}}(2)$ algebra generated by the highest vector $v$ such that:
\begin{align*}
e_nv=0,\,\, \text{for $n\geq0$};\qquad f_nv=h_nv=0,\,\, \text{for $n>0$};\qquad h_0v=hv,\,\, Kv=kv.
\end{align*}
We denote by $\Asl(2)_k$ the quotient of the universal enveloping algebra, which acts on the integrable representations $\calL_{h,k}$ (the representations of the level $k$).  In this section we consider only level 1 representations.

Each integrable representation of $\Asl(2)$ has two natural grading. The
first of them is a $h_0$ grading. Another grading is defined by Sugawara operator $L_0$ given by
\begin{align*}
L_0^{(\Asl)}=\frac{1}{2(k+2)}\left(\sum_{k>0}\left(2e_{-k}f_k+2f_{-k}e_k+h_{-k}h_k\right)+e_0f_0+f_0e_0+\frac{h_0^2}{2}\right).
\end{align*}
This operator has very simple properties on the representations $\calL_{h,k}$:
\begin{align*}
[L_0^{(\Asl)},e_{-n}]=ne_{-n}, \quad [L_0^{(\Asl)},h_{-n}]=nh_{-n}, \quad [L_0^{(\Asl)},f_{-n}]=nf_{-n}, \quad L_0^{(\Asl)}v=\frac{h(h+2)}{4(k+2)}.
\end{align*}
We refer to such grading as a \emph{homogeneous grading}. The character of the representation $\calL_{h,k}$ defined by the formula
\begin{align*}
\chi_{\Asl}^{h,k}=\left.\text{Tr}\,q^{L_0}t^{h_0/2}\right|_{\calL_{h,k}}.
\end{align*}

\secpart The level 1 representations $\calL_{0,1}$ and $\calL_{1,1}$ have the simple construction due to Frenkel and Kac \cite{Frenkel Kac} in terms of the one Heisenberg algebra $h_n$ and the operator $D$ . Denote by $\rmF_{P}$ the Fock representation of the Heisenberg algebra $\langle h_n \rangle$ with the vacuum vector $v_P$:
\begin{align*}
h_nv_P=0\;\, \text{for $n>0$},\quad h_0v_P=Pv_P.
\end{align*}
Denote by $D\colon \rmF_P \rightarrow \rmF_{P+1}$ the operator defined by the commutation relations
\begin{align*}
[h_n,D]=0\;\,\text{for $n\ne 0$},\quad [h_0,D]=D.
\end{align*}
Then $\widehat{\mathfrak{sl}}(2)_1$ representations $\mathcal{L}_{0,1}$ and $\mathcal{L}_{1,1}$ can be realized as the direct sums of the Fock modules, namely
\begin{align}\label{eq:L1:F}
\mathcal{L}_{0,1}=\bigoplus_{n \in \mathbb{Z}} \rmF_{2n},\qquad \mathcal{L}_{1,1}=\bigoplus_{n \in \mathbb{Z}} \rmF_{2n+1}.
\end{align}
The generators $e_i$ and $f_i$ are defined in both representations by the formulae
\begin{align*}
e(z)= \sum_{n \in \mathbb{Z}} e_nz^{-n}=z^{h_0/2}D^2\exp\Bigl(2\sum_{n \in \mathbb{Z}_{>0}} \dfrac{h_{-n}}{2n} z^{n} \Bigr) \exp\Bigl(2\sum_{n \in \mathbb{Z}_{>0}} \dfrac{h_n}{-2n} z^{-n} \Bigr)z^{h_0/2},\\ f(z)=\sum_{n \in \mathbb{Z}} f_nz^{-n}=z^{h_0/2}D^2\exp\Bigl(-2\sum_{n \in \mathbb{Z}_{>0}} \dfrac{h_{-n}}{2n} z^{n} \Bigr) \exp\Bigl(-2\sum_{n \in \mathbb{Z}_{>0}} \dfrac{h_n}{-2n} z^{-n} \Bigr)z^{h_0/2}.\end{align*}
It is convenient to consider operators $h_n$ as modes of the bosonic field $\varphi(z)$:
\begin{align} \label{eq:varphi}
\varphi(z)=\sum_{n \in \mathbb{Z}\setminus 0} \dfrac{h_n}{-2n} z^{-n}+\frac{h_0\log z}2+\widehat{Q},
\end{align}
where the operator $\widehat{Q}$ is conjugate to the operator $\widehat{P}=h_0$, i.e. defined by the relation: $[h_0,\widehat{Q}]=1$. Therefore $D=\exp{\widehat{Q}}$. Using the field  $\varphi(z)$ the free field realization of $\Asl(2)_1$ can be simply rewritten~as:
\begin{align}\label{eqFrenkel_Kac}
\sum_{n \in \mathbb{Z}}\! e_nz^{-n}=:\!\exp\Bigl(2 \varphi(z) \Bigr)\!:,\quad
\sum_{n \in \mathbb{Z}}\! f_nz^{-n}=:\!\exp\Bigl(-2\varphi(z) \Bigr)\!:,\quad \sum_{n \in \mathbb{Z}}\! h_nz^{-n}=2z \partial_z \varphi(z).
\end{align}
where $:\ldots:$ denotes the creation-annihilation normal ordering.

The formulas for the characters of the representations $\calL_{0,1}$ and $\calL_{1,1}$ follow from the  construction \eqref{eq:L1:F}:
\begin{align}
\chi_{\Asl}^{0,1}=\sum_{n \in \Z} t^n q^{n^2} \chi_B(q), \;\; \chi_{\Asl}^{1,1}=\sum_{r\in \Z+\frac12} t^r q^{r^2} \chi_B(q), \qquad \text{where}\;\; \chi_B(q)=\prod_{k \in \Z_{>0}} \frac1{1-q^k}. \label{eqchi_Asl1}
\end{align}

\secpart We want to construct the special basis in the representation of the algebra $\calA(1,2)=\calH\oplus \Asl(2)_1$. The basic vectors are labeled by the torus fixed points on the moduli space $\bigsqcup_N\calM(1,N)^{\Z_2}$. In this case the torus fixed points are labeled by two colors colored Young diagrams $\lambda$ with the angle colored in the  color $\sigma$ (see Appendix \ref{AppendixYoung} for the notation on colored partitions). We will denote the corresponding basic vector by a $J_{\lambda^\sigma}$. The color of the angle $\sigma$ corresponds to the highest weight of representations $\calL_{\sigma,1}$.

It follows, from the definition of the geometric action of operators $e_{-1},f_0,e_0,f_1$ (see \cite{Nakajima:1998}) that combinatorial characteristics agree with the algebraic grading as follows:
\begin{align*}
h_0(J_{\lambda^\sigma})=(2d(\lambda^\sigma)+\sigma)J_{\lambda^\sigma}, \qquad L_0(J_{\lambda^\sigma})=\frac{2|\lambda|+h_0}4J_{\lambda^\sigma},
\end{align*}
where $d(\lambda^\sigma)=N_0(\lambda^\sigma)-N_1(\lambda^\sigma)$ is the difference between the number of white and black boxes in diagram $\lambda^\sigma$, $L_0$ is a total degree with respect to the algebra $\calH\oplus\Asl(2)_1$:
\begin{align}\label{eq:L0_12}
L_0=L_0^{\Asl}+L_0^{H}=L_0^{\Asl}+\frac12\sum_{k \in \Z_{>0}} w_{-k}w_k.
\end{align}
The existence of such a basis means that we can compute the character of the representation of the algebra $\calA(1,2)$. Using the generating functions $\chi^{(1)}_{d,\sigma}$ introduced in the Appendix \ref{AppendixYoung} and the formula \eqref{eqchi^(1)} we get:
\begin{align*}
&\chi_\sigma^{(1)}:=\sum_{\lambda^\sigma} q^\frac{|\lambda|+d(\lambda^\sigma)+\sigma/2}2 t^{d(\lambda^\sigma)+\sigma/2}=\sum_{d}  (t\sqrt{q})^{d+\sigma/2} \chi^{(1)}_{d,\sigma}(q)= \\
&= \sum_{d}  (t\sqrt{q})^{d+\sigma/2} q^{\frac{2d^2-(-1)^\sigma d}2} \chi_B(q)^2
=\sum_{n \in \Z +\frac{\sigma}2} t^nq^{n^2} \chi_B(q)^2.
\end{align*}

Due to the formulas \eqref{eqchi_Asl1} the last expressions coincide with the characters of $\calH\oplus\Asl(2)_1$ representations:
\begin{align*}
\calF \otimes \calL_{0,1}=\bigoplus_{n \in \mathbb{Z}} \calF\otimes\rmF_{2n},\qquad \calF \otimes \calL_{1,1}=\bigoplus_{n \in \mathbb{Z}} \calF\otimes\rmF_{2n+1},
\end{align*}
where $\calF$ is a Fock representation of $\mathcal{H}$. Each vector in these representations can be obtained from the vacuum vector by action of $w_n,h_n,D$ so it is natural to ask for the expression of $J_{\lambda^\sigma}$ in terms of these generators.

\subsection{The limit and principal realization}
We will study the algebra $\calA(1,2)$ using the limit of the quantum toroidal $\mathfrak{gl}_1$ algebra $\calE_1(q,t)$ depending on two parameters $q,t$.

\secpart We will mainly follow \cite{AFHKSY:2011} in this and next subsubsections.

The {\it quantum toroidal $\mathfrak{gl}_1$} algebra
is an associative algebra $U(q,t)$ is generated by the $E_i, F_i$, $i \in \Z$, $\psi^+_k, \psi^-_{-k}$, $k\in \Z_{\geq 0}$ and the central element $C$. The relations are written in terms of the currents
\begin{align*}
E(z)=\sum_{n\in \Z} E_n z^{-n}, \quad F(z)=\sum_{n\in \Z} F_n z^{-n}, \quad
\psi^\pm(z)=\sum_{\pm n\in \Z_{\ge0}}\psi^\pm_n z^{-n}
\end{align*}
and have the form:
\begin{align*}
&\psi^\pm(z) \psi^\pm(w) = \psi^\pm(w) \psi^\pm(z),
 \qquad\qquad
 \psi^+(z)\psi^-(w) =
 \dfrac{g(C^2 w/z)}{g(C^{-2}w/z)}\psi^-(w)\psi^+(z),
\\
&\psi^+(z)E(w) = g(C^{-1}w/z)^{-1} E(w)\psi^+(z),
 \qquad
 \psi^-(z)E(w) = g(C^{-1}z/w) E(w)\psi^-(z),
\\
&\psi^+(z)F(w) = g(Cw/z) F(w)\psi^+(z),
 \qquad
 \psi^-(z)F(w) = g(Cz/w)^{-1} F(w)\psi^-(z),
\\
&[E(z),F(w)]
 =\dfrac{(1-q)(1-1/t)}{1-q/t}
 \big( \delta(C^{-2}z/w)\psi^+(C w)-
 \delta(C^2 z/w)\psi^-(C^{-1}w) \big),
\\
&G^{-1}(z/w)E(z)E(w)=G(z/w)E(w)E(z), \qquad G(z/w)F(z)F(w)=G^{-1}(z/w)F(w)F(z), \\
& [E_0,[E_1,E_{-1}]]=[F_0,[F_1,F_{-1}]]=0,
\end{align*}
where
\begin{align*}
\delta(z)=\sum_{n\in\Z} z^n, \qquad g(z)\seteq\dfrac{G^+(z)}{G^-(z)},\quad
G^\pm(z)\seteq(1-q^{\pm1}z)(1-t^{\mp 1}z)(1-q^{\mp1}t^{\pm 1}z).
\end{align*}

Note that the elements $\psi^+_0, \psi^-_0$ are central.
The representation $V$ is said to be \emph{of level $(r,l)$} if
\begin{align*}
C v = (t/q)^{r/4}v, \qquad (\psi^-_0/\psi^+_0)v=(t/q)^{l}v
\end{align*}
for any $v \in V$. In this paper we consider only the $l=0$ case so we will denote level just by $r$ for simplicity.

It was proven in \cite{Feigin_Shiraishi:2009} that there exists the infinite  system of commuting elements $I_1,I_2, \dots $ in $\calE_1$ (system of Integrals of motion). The first of them has the form $I_1=E_0.$
%\begin{equation}\label{eqIntegrals_DI}
%I_1=E_0,\quad I_2=[E_{-1},E_1],\quad I_3=[E_{-1},[E_0,E_1]], \quad I_3=[E_{-1},[E_0,[E_0,E_1]]]
%\end{equation}

\secpart The level 1 realization of $\calE_1(q,t)$ is given in the Fock representation of the (deformed) Heisenberg algebra.  Consider the algebra with the generators $a_n$ ($n \in \Z\setminus\{0\}$) and the relations:
\begin{align}\label{eqHeisenberg_DI}
[a_{n},a_{m}] = n \frac{1-q^{|n|}}{1-t^{|n|}}\delta_{n+m}.
\end{align}
Denote by $\calF$ the Fock representation of this Heisenberg algebra generated by the vacuum vector $|0\ra$ such that
\begin{align*} a_n|0\ra=0\quad \text{for $n>0$}.
\end{align*}
It was proven in \cite{Feigin_Shiraishi:2009} that for any $u \in \mathbb{C}$ the following formulae give the level 1 representation of the algebra $\calE_1$
\begin{equation} \label{eqDI_Fock}
\begin{aligned}
& C \mapsto (t/q)^{1/4}  \\
&E(z)\mapsto
u\exp\Big( \sum_{n=1}^{\infty} \dfrac{1-t^{-n}}{n}a_{-n} z^{n} \Big)
\exp\Big(-\sum_{n=1}^{\infty} \dfrac{1-t^{n} }{n}a_n    z^{-n}\Big),\\
&F(z)\mapsto u^{-1}
\exp\Big(-\sum_{n=1}^{\infty} \dfrac{1-t^{-n}}{n}(t/q)^{n/2}a_{-n} z^{n}\Big)
\exp\Big( \sum_{n=1}^{\infty} \dfrac{1-t^{n}}{n} (t/q)^{n/2} a_n z^{-n}\Big),\\
&\psi^+(z)\mapsto
\exp\Big(
 -\sum_{n=1}^{\infty} \dfrac{1-t^{n}}{n} (1-t^n q^{-n})(t/q)^{-n/4} a_n z^{-n}
    \Big), \\
&\psi^-(z)\mapsto
\exp\Big( \sum_{n=1}^{\infty} \dfrac{1-t^{-n}}{n} (1-t^n q^{-n})(t/q)^{-n/4} a_{-n}z^{n}
    \Big).
\end{aligned}
\end{equation}
We denote this representation by $\calF_u$.

\secpart Consider the limit
\begin{align}
q=-e^{-\tau \eps_{2}},\; t=-e^{\tau \eps_{1}}, \qquad \tau \rightarrow 0,\; \eps_1=b,\; \eps_2=b^{-1}. \label{eq:qt:limit}
\end{align}
The limit of the Heisenberg algebra \eqref{eqHeisenberg_DI} has the relations:
\begin{align*}
[a_{n},a_{m}] =  \begin{cases}
-nb^{-2}\delta_{n+m,0}, \quad \;
&\textrm{if} \quad  n \equiv  0 \;\textrm{mod}\; 2
 \\ \;\;\, n \delta_{n+m,0}, \quad &\textrm{if} \quad  n \equiv 1 \;\textrm{mod}\; 2 \end{cases}
\end{align*}
We see that the Heisenberg algebra falls into two pieces: the even and the odd part. For simplicity we make the replacement $a_{2n} \mapsto ib^{-1}a_{2n}$ and get the standard Heisenberg algebra
\begin{align*}
[a_{n},a_{m}] =\begin{cases}
n\delta_{n+m,0}, \quad \textrm{if} \quad  n \equiv  0 \;\textrm{mod}\; 2 \\
n \delta_{n+m,0}, \quad \textrm{if} \quad  n \equiv 1 \;\textrm{mod}\; 2 \end{cases}
\end{align*}

The $u$ (parameter of the representation $\calF_u$) in the limit reads:
$u=(-1)^\sigma e^{\kappa\tau}$, where $\sigma\in\{0,1\}$ is a discrete parameter and $\kappa\in \mathbb{C}$ is a continuous parameter. We denote the limit representation by $\calF^{(\sigma)}(\kappa)$. The limit of generators in $\calF^{(\sigma)}(\kappa)$ reads:
\begin{align}
&E(z), F(z) \mapsto
:(-1)^\sigma\exp\left(2\sum \frac{a_{2n+1}}{-(2n+1)} z^{-(2n+1)} \right):+O(\tau), \notag \\
&\frac{\psi^\pm(z)-\psi^\pm(-z)}2 \mapsto \mp \left( 2Q\sum\nolimits_{\pm n \in \Z_{>0}} a_{2n+1}    z^{-(2n+1)}\right) \tau +O(\tau^2),\label{eqr=1_limit} \\
&\frac{\psi^\pm(z)+\psi^\pm(-z)}2 \mapsto 1+ \left(2Q^2(\sum_{\pm n \in \Z_{>0}} a_{2n+1}    z^{-(2n+1)})^2-2Q\epsilon_1 \sum_{\pm n \in \Z_{>0}} 2n a_{2n} z^{-2n}\right)\tau^2+O(\tau^3). \notag
\end{align}
The formula for the limit of $E(z)$ recovers the principal (Lepowsky--Wilson) realization of $\widehat{\mathfrak{sl}}(2)_1$ which we recall in the next paragraph.

\secpart One can introduce another set of generators in the algebra $\Asl(2)$:
\begin{align}
a_{2n+1}=f_{n+1}+e_n,\quad b_{2n+1}=f_{n+1}-e_n, \quad b_{2n}=h_n-\frac12\delta_{n,0}K. \label{eqGenLW}
\end{align}
This generators satisfy the relations
\begin{align*}
&[a_{2n+1},a_{2m+1}]=(2n+1)\delta_{n+m+1},\quad [a_{2n+1},b_m]=2b_{m+2n+1}, \quad [b_{2n},b_{2m}]=2n\delta_{m+n} \\ &[b_{2n+1},b_{2m+1}]=-(2n+1)\delta_{n+m+1},\quad [b_{2n+1},b_{2m}]=2a_{2m+2n+1}.
\end{align*}
It was proven by Lepowsky and Wilson \cite{Lepowsky_Wilson} that the integrable representations $\mathcal{L}_{\sigma,1}$ (where $\sigma=0,1$) can be realized as the Fock module $\mathbb{C}[a_{-1},a_{-3},a_{-5},\ldots]$ over Heisenberg algebra $a_{2n+1}$. The action of the generators $b_n$ on this representation is defined by the formula
\begin{align}
\sum_n b_nz^{-n}=b(z)=\frac{(-1)^{\sigma+1}}2\exp\left(2\sum_{n} \dfrac{a_{2n+1}}{-2n-1} z^{-2n-1} \right). \label{eqLepovsky_Wilson}
\end{align}
Therefore, in the limit $\calE_1(q,t)$ operators $E_n$ and $\psi^\pm_{2n+1}$ give generators of $\Asl(2)$: $b_n$ and $a_{2n+1}$ correspondingly. The remaining set of generators $\psi^{\pm}_{2n}$ converges to the additional Heisenberg algebra $w_n$. If we denote the limit of the representation $\calF_u$, $u=(-1)^\sigma e^{\kappa\tau}$, $\tau \rightarrow 0$ by $\calF^{(\sigma)}(\kappa)$ then we get:
\begin{align*}
\calF^{(\sigma)}(\kappa)=\calF\otimes \calL_{\sigma,1},
\end{align*}
where $\calF$ is a Fock representation of $\mathcal{H}$. The parameter $\kappa$ doesn't appear in the formulas for the generators but appear in the following formulas for the vertex operator.

We recapitulate part of this subsection as follows: there exist two bosonizations of the algebra $\calA(1,2)$. The first of them is based on the  homogeneous realization and has the generators $w_n,h_n,D$. The second is based on the principal realization and has the generators $a_n$.  Due to \eqref{eqFrenkel_Kac} and \eqref{eqLepovsky_Wilson}  the relation between these two constructions reads:
\begin{align}\label{eq:h,w:a}
&w_n=a_{2n}, \quad  z\partial \varphi(z^2)=\sum h_nz^{-2n}=\frac12+ \frac{(-1)^{\sigma+1}}4\Bigl(\exp(2\phi)+\exp(-2\phi)\Bigr)
\end{align}
and
\begin{align*}
&a_{2n}=w_n, \quad z\partial \phi(z^2)=\sum a_{2n+1}z^{-2n-1}=  z^{-1}\exp\Bigl(2\varphi(z^2)\Bigr)+ z\exp\Bigl(-2\varphi(z^2)\Bigr),
\end{align*}
where we used field notations: $\varphi(z)$ introduced in  \eqref{eq:varphi} and
\begin{align}\label{eq:phi}
\phi(z)=\sum_{n\in \Z} \dfrac{a_{2n+1}}{-2n-1} z^{-2n-1}.
\end{align}

\subsection{Basis}

In this subsection we will construct the basis in the representations of the algebra $\calA(1,2)=\calH\oplus\Asl(2)_1$ using the limit of the algebra $\calE_1(q,t)$  described above.

\secpart Denote by $\iota$ the isomorphism between the Fock space $\mathcal{F}_u$ and the space of symmetric polynomials $\Lambda$:
 \begin{align*}
\iota: \mathcal{F}_u \simto \Lambda,\;\;
       a_{\lambda} \mapsto p_\lambda,\qquad
\text{where}\;\;\;
p_\lambda=p_{\lambda_1}\cdot p_{\lambda_2}\cdot\ldots\cdot p_{\lambda_k},\;\; p_n=\sum x_i^n.
\end{align*}
This isomorphism converts Shapovalov form on $\calF_u$ to the Macdonald scalar product on~$\Lambda$~\eqref{eqMacdonald_produc}. The Macdonald polynomials $J_\lambda(q,t)$ (integral form, see Appendix \ref{Appendix_Macdonald_polyn}) form an orthogonal basis in $\Lambda$. Denote their preimages by the same letter $J_\lambda$. First examples have the form:
\begin{align*}
&J_{{\varnothing}}=|u\ra, \qquad J_{(1)}= (1-t)a_{-1}|u\ra, \\
&J_{(2)}=\left(\frac{1}{2}(1+q)(1-t)^2a_{-1}^{2} +
\frac{1}{2}(1-q)(1-t^2)a_{-2}\right)|u\ra,\\
&J_{(1,1)}=\left(\frac{1}{2}(1-t)^2(1+t)(a_{-1}^{2}-a_{-2})\right)|u\ra.
\end{align*}
 The operator $I_1=E_0$ under isomorphism $\iota$ is identified with the Macdonald difference operator. This operator is diagonal on the basis $J_\lambda$
\begin{align} \label{eqDI_I1}
E_0 |J_\lambda\rangle
=u \varepsilon_\lambda|J_\lambda\rangle,\quad
\varepsilon_\lambda\seteq 1+(t-1) \sum (q^{\lambda_j}-1)t^{-j}.
\end{align}

\secpart The Macdonald polynomials $J_{\lambda}^{q,t}$ have a well defined limit when $q,t \rightarrow -1$ \cite{Uglov 1}. We denote this limit by the \emph{rank 2 Uglov polynomials} $J_{\lambda}^{(-1/b^2,2)}$. Many properties of these polynomials simply follow from the properties of the Macdonald polynomials, we collect them in Appendix~\ref{Appendix_Uglov_polyn}.  It is convenient to introduce $J^{(2)}_\lambda$ by:
\begin{align} \label{eq:J(2):def}
J^{(2)}_\lambda=\lim_{\tau \rightarrow \infty}\left(\frac{ (-1)^{n(\lambda)}}{\tau^{|\lambda^\lozenge|}\, 2^{|\lambda|-|\lambda^\lozenge|}}\,J_\lambda(q,t)\right).
\end{align}
First examples have the form:
\begin{align}\label{eq:J(2):exmp}
&J_{{\varnothing}}^{(2)}=|\kappa\ra^{\sigma}, \qquad J_{(1)}^{(2)}= a_{-1}|\kappa\ra^{\sigma}, \qquad J_{(2)}^{(2)}= (b^{-1} a_{-1}^2-i a_{-2})|\kappa\ra^{\sigma},\notag \\
&J_{(1,1)}^{(2)}=(b a_{-1}^2-i a_{-2})|\kappa\ra^{\sigma}, \qquad J_{(3)}^{(2)}=(\frac{1}{3 b}a_{-1}^3-i a_{-2}a_{-1}+\frac{2}{3 b} a_{-3})|\kappa\ra^{\sigma}, \\
&J_{(2,1)}^{(2)}= -\frac{1}{3}(a_{-1}^3-a_{-3})|\kappa\ra^{\sigma}, \qquad J_{(1,1,1)}^{(2)}=(\frac{b}{3}a_{-1}^3-ia_{-2}a_{-1}+\frac{2b}{3}a_{-3})|\kappa\ra^{\sigma},\notag \\
&J_{(2,2)}^{(2)}=(-iQa_{-4}+\frac{4}{3}a_{-3}a_{-1}-a_{-2}^2-\frac{1}{3} a_{-1}^4)|\kappa\ra^{\sigma}. \notag
\end{align}
Here and below $|\kappa\ra^{\sigma}$ denotes the vacuum vector in the representation $\calF^{(\sigma)}(\kappa)$. We assign to each vector $J^{(2)}_\lambda \in \calF^{(\sigma)}(\kappa)$ the Young diagram $\lambda$ colored in two colors with the angle of color $\sigma$ (see Appendix \ref{AppendixYoung} for the definition and notation $d(\lambda)$, $\lambda^\lozenge$).

From the orthogonality of the Macdonald polynomials follows that $J^{(2)}_\lambda|\kappa\ra^{\sigma}$ are orthogonal with norms given by (see \eqref{eqNorm_Uglov}):
\begin{align}\label{eqNorm_r1}
\langle J_{\lambda}^{(2)} |J_{\mu}^{(2)} \rangle
=\delta_{\lambda, \mu} (-1)^{|\lambda^\lozenge|}\prod_{s \in \lambda^{\lozenge}}\left(bl_\lambda(s)+b- \frac{a_\lambda(s)}{b}\right)\left(-b l_\lambda(s)+\frac{a_\lambda(s)+1}{b}\right).
\end{align}

The number of boxes $|\lambda|$ equals to the degree of the Uglov polynomial $J_\lambda^{(2)}$. The action of operators $a_{-n}, b_{-n}$ increases this number by $n$. Hence $|\lambda|$ coincides with the the principal grading $\mathrm{Pr}$ for the algebra $\calH\oplus\Asl(2)_1$ i.e. creation operators $e_{-1}$ and $f_0$ increase this grading by 1.

\secpart In the limit of the commutative subalgebra in the algebra $\calE_1(q,t)$ one can find the commutative subalgebra (the system of the Integrals of Motion) which act diagonally on the basis $J_\lambda^{(2)}$. For example from formula \eqref{eqDI_I1} follows that
\begin{align}
E_0 \rightarrow 1-2h_0, \qquad h_0(J_{\lambda^\sigma}^{(2)})= \left(2d(\lambda)+\sigma\right)J_{\lambda^\sigma}^{(2)}. \label{eq:r1:h0}
\end{align}
We add superscript $\sigma$ (color of the angle of $\lambda^\sigma$ or the highest weight of $\calL_{\sigma,1}$) in notation where formulas depend on it. The formula for action of $h_0$ \eqref{eq:r1:h0} is equivalent to the fact that the vectors
\begin{align}\label{eqJ_basis_F0}
&J^{(2)}_{\lambda^0},\,  d(\lambda)=d\quad \text{form a basis in}\quad \calF\otimes\mathrm{F}_{2d}, \\ \label{eqJ_basis_F1}
&J^{(2)}_{\lambda^1},\,  d(\lambda)=d\quad \text{form a basis in}\quad \calF\otimes\mathrm{F}_{2d+1}.
\end{align}
The vectors $J^{(2)}_{\lambda^\sigma}$ with $|\lambda|$ minimal for given $d(\lambda)$ are the highest vectors   in $\calF\otimes\mathrm{F}_{d}$. Such $\lambda$ has triangular form and called 2-core (see Appendix~\ref{AppendixYoung}).

The homogeneous grading defined by operator $L_0$ \eqref{eq:L0_12} is determined by the principal grading $\mathrm{Pr}$ and the $h_0$ grading by formula:
\[L_0=\frac{2\mathrm{Pr}+h_0}4,\quad \text{i.e.} \qquad L_0(J_{\lambda^\sigma}^{(2)})= \left(\frac{2|\lambda|+2d(\lambda)+\sigma}4\right)J_{\lambda^\sigma}^{(2)}.\]

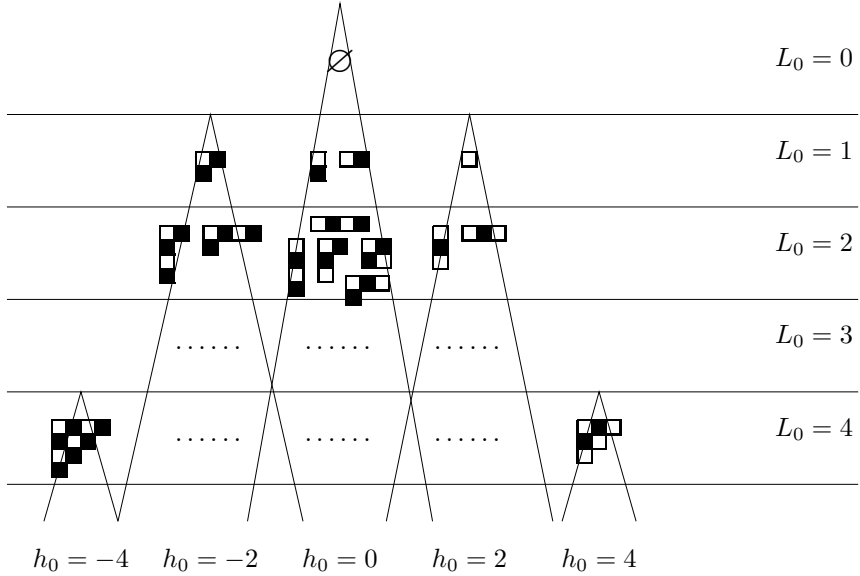
\begin{figure}[h]
\ytableausetup{mathmode, boxsize=0.5em}
\begin{center}
\begin{tikzpicture}[x=0.7em, y=0.7em]
\draw (0,-1) node[anchor=north]
	{\Large $\varnothing$};
\draw (7,-6.5) node[anchor=north]
	{\begin{ytableau}
	*(white)
	\end{ytableau}};
\draw (0,-6.5) node[anchor=north]
	{\begin{ytableau}
	*(white) \\ *(black)
	\end{ytableau}\;
	\begin{ytableau}
	*(white) & *(black)
	\end{ytableau}};
\draw (-7,-6.5) node[anchor=north]
	{\begin{ytableau}
	*(white)& *(black)\\
	*(black)
	\end{ytableau}};
\draw (7,-10.5) node[anchor=north]
	{\begin{ytableau}
	*(white) \\ *(black) \\ *(white)
	\end{ytableau}\;
	\begin{ytableau}
	*(white) & *(black) & *(white)
	\end{ytableau}};
\draw (0,-10) node[anchor=north]
	{\begin{ytableau}
	*(white) & *(black) & *(white) & *(black)
	\end{ytableau}};
\draw (0,-11.2) node[anchor=north]
	{\begin{ytableau}
	*(white) \\ *(black) \\ *(white) \\ *(black)
	\end{ytableau}\;
	\begin{ytableau}
	*(white) & *(black) \\ *(black) \\ *(white)
	\end{ytableau}\;
	\begin{ytableau}
	*(white) & *(black) \\ *(black) & *(white)
	\end{ytableau}};
\draw (1.5,-13.2) node[anchor=north]	
	{\begin{ytableau}
	*(white) & *(black) & *(white) \\ *(black)
	\end{ytableau}};
\draw (-7,-10.5) node[anchor=north]
	{\begin{ytableau}
	*(white) & *(black) \\ *(black) \\ *(white) \\ *(black)
	\end{ytableau}\;
	\begin{ytableau}
	*(white) & *(black) & *(white) & *(black) \\ *(black)
	\end{ytableau}};
\draw (14,-21) node[anchor=north]
	{\begin{ytableau}
	*(white) & *(black) & *(white) \\
	*(black) & *(white) \\
	*(white)
	\end{ytableau}};
\draw (-14,-21) node[anchor=north]
	{\begin{ytableau}
	*(white) & *(black) & *(white) & *(black)\\
	*(black) & *(white) & *(black)\\
	*(white) & *(black)\\
	*(black)
	\end{ytableau}};
\foreach \x in {-1,0,1}
	{ 
		\draw (7*\x,-17) node[anchor=north] {\dots\dots};
		\draw (7*\x,-22) node[anchor=north] {\dots\dots};
	}
%Vertical angles
\draw[very thin] (-12,-27) -- (-14,-20) -- (-16,-27);
\draw[very thin] (-2,-27) -- (-7,-5) -- (-12,-27);
\draw[very thin] (-5,-27) -- (0,1) -- (5,-27);
\draw[very thin] (2.5,-27) -- (7,-5) -- (11.5,-27);
\draw[very thin] (12,-27) -- (14,-20) -- (16,-27);
%Horizontal lines and values of L_0
\foreach \x in {0,1,2,3,4}
	{\draw (23,-2-5*\x) node[anchor=west] {$L_0=\x$};
	\draw[very thin] (-18, -5-5*\x) -- (28, -5-5*\x);
	};
\foreach \x in {-4,-2,0,2,4}
	\draw (3.5*\x,-28) node[anchor=north] {$h_0=\x$};
\end{tikzpicture}
\caption{\small{The basis in $\calF^{(0)}(\kappa)$.  Colored diagram $\lambda$ represents a vector  $J^{(2)}_{\lambda^0}$.
The interior of each angle corresponds to the representation $\calF \otimes \rmF_{2d} \subset \calF^{(0)}(\kappa)$.}} \label{fig_01}
\end{center}
\end{figure}
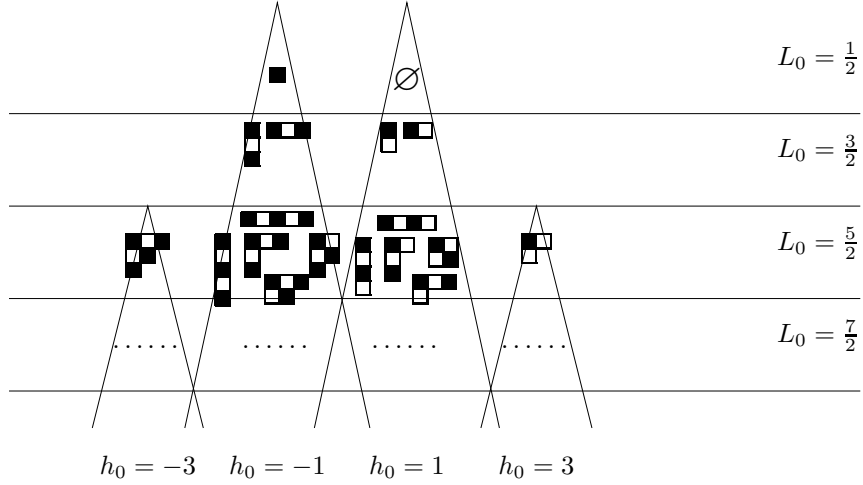
\begin{figure}[h]
\ytableausetup{mathmode, boxsize=0.5em}
\begin{center}
\begin{tikzpicture}[x=0.7em, y=0.7em]
\draw (3.5,-2) node[anchor=north]
	{\Large $\varnothing$};
\draw (-3.5,-2) node[anchor=north]
	{\begin{ytableau}
	*(black)
	\end{ytableau}};
\draw (3.5,-5) node[anchor=north]
	{\begin{ytableau}
	*(black) \\ *(white)
	\end{ytableau}
	\begin{ytableau}
	*(black) & *(white)
	\end{ytableau}} ;
\draw (-3.5,-5) node[anchor=north]
	{\begin{ytableau}
	*(black) \\ *(white) \\ *(black)
	\end{ytableau}
	\begin{ytableau}
	*(black) & *(white) & *(black)
	\end{ytableau}};
\draw (10.5,-11) node[anchor=north]
	{\begin{ytableau}
	*(black) & *(white) \\ *(white)
	\end{ytableau}} ;
\draw (3.5,-10) node[anchor=north]
	{\begin{ytableau}
	*(black) & *(white) & *(black) & *(white)
	\end{ytableau}};
\draw (3.5,-11.2) node[anchor=north]
	{\begin{ytableau}
	*(black) \\ *(white) \\ *(black) \\ *(white)
	\end{ytableau}\;
	\begin{ytableau}
	*(black) & *(white) \\ *(white) \\ *(black)
	\end{ytableau}\;
	\begin{ytableau}
	*(black) & *(white)  \\ *(white) & *(black)
	\end{ytableau}};
\draw (5,-13.2) node[anchor=north]	
	{\begin{ytableau}
	*(black) & *(white) & *(black) \\ *(white)
	\end{ytableau}};
\draw (-3.5,-9.8) node[anchor=north]
	{\begin{ytableau}
	*(black) & *(white) & *(black) & *(white) & *(black)
	\end{ytableau}};
\draw (-3.5,-11) node[anchor=north]
	{\begin{ytableau}
	*(black) \\ *(white) \\ *(black) \\ *(white) \\ *(black)
	\end{ytableau}\;
	\begin{ytableau}
	*(black) & *(white) & *(black) \\ *(white) \\ *(black)
	\end{ytableau}\;\;
	\begin{ytableau}
	*(black) & *(white)  \\ *(white) & *(black) \\ *(black)
	\end{ytableau}};
\draw (-3,-13.2) node[anchor=north]	
	{\begin{ytableau}
	*(black) & *(white) & *(black) \\ *(white) & *(black)
	\end{ytableau}};
\draw (-10.5,-11) node[anchor=north]
	{\begin{ytableau}
	*(black) & *(white) & *(black)\\
	*(white) & *(black) \\
	*(black)
	\end{ytableau}};
\foreach \x in {-3,-1,1,3}
	\draw (3.5*\x,-17) node[anchor=north] {\dots\dots};
%Vertical angles
\draw[very thin] (-13.5,-22) -- (-10.5,-10) -- (-7.5,-22);
\draw[very thin] (-8.5,-22) -- (-3.5,1) -- (1.5,-22);
\draw[very thin] (-1.5,-22) -- (3.5,1) -- (8.5,-22);
\draw[very thin] (13.5,-22) -- (10.5,-10) -- (7.5,-22);
%%Horizontal lines and values of L_0
\foreach \x/\xtext in {0/\frac12,1/\frac32,2/\frac52,3/\frac72}
	{\draw (23,-2-5*\x) node[anchor=west] {$L_0=\xtext$};
	\draw[very thin] (-18, -5-5*\x) -- (28, -5-5*\x);
	};
\foreach \x in {-3,-1,1,3}
	\draw (3.5*\x,-23) node[anchor=north] {$h_0=\x$};
\end{tikzpicture}
\caption{\small{The basis in $\calF^{(1)}(\kappa)$.} \label{fig_11} }
\end{center}
\end{figure}
\secpart It is convenient to rewrite the basis $J^{(2)}_{\lambda^\sigma}$ in terms of the homogeneous bosonization (with generators $w_n,h_n,D$). First note that $D$ commutes with $h_n,w_n$ for $n \neq 0$. Then action of $D$ translates the highest vector of $\calF \otimes \rmF_{d}$ to the highest vector of $\calF \otimes \rmF_{d+1}$. From \eqref{eqNorm_r1} follows that norms $\la J^{(2)}_{\lambda^\sigma}|J^{(2)}_{\lambda^\sigma}\ra=1$ if $\lambda$ is a 2-core. Hence all such vectors can be obtained by the action of $D$ to the vacuum $|\kappa\ra^\sigma$:
\begin{align*}
&J_{\varnothing^0}^{(2)}=|\kappa\ra^0, \quad J_{(1)^0}^{(2)}=D^2|\kappa\ra^0, \quad J_{(2,1)^0}^{(2)}=D^{-2}|\kappa\ra^0, \quad J_{(3,2,1)^0}^{(2)}=D^4|\kappa\ra^0, \ldots \\
&J_{\varnothing^1}^{(2)}=|\kappa\ra^1, \quad J_{(1)^1}^{(2)}=D^{-2}|\kappa\ra^1, \quad J_{(2,1)^1}^{(2)}=D^{2}|\kappa\ra^1, \quad J_{(3,2,1)^1}^{(2)}=D^{-4}|\kappa\ra^1, \ldots
\end{align*}
Any other $J^{(2)}_{\lambda^\sigma}$  can be obtained from $J^{(2)}_{\widetilde{\lambda}^\sigma}$ by action of generators of the two Heisenberg algebras $h_n\in \widehat{\mathfrak{sl}}(2)$ and $w_n\in \calH$, where $\widetilde{\lambda}$ is $2$--core of partition $\lambda$. The first examples in the space $\calF\otimes\mathrm{F}_{0}$ are:
\begin{align*}
&J_{\varnothing^0}^{(2)}=|\kappa\ra^0 \quad J_{(2)^0}^{(2)}=-(i w_{-1}+b^{-1}h_{-1})|\kappa\ra^0, \quad J_{(1,1)^0}^{(2)}= -(i w_{-1}+bh_{-1})|\kappa\ra^0,\notag\\
&J_{(1,1,1,1)^0}^{(2)}=(-2i b w_{-2}-w_{-1}^2+ 2i b w_{-1}h_{-1}+bh_{-1}^2-2bh_{-2})|\kappa\ra^0, \notag \\
&J_{(2,1,1)^0}^{(2)}=\left(-2ibw_{-2}-w_{-1}^2+i(b^{-1}-b)w_{-1}h_{-1}+b^2h_{-1}^2-(1-b^2) h_{-2}\right)|\kappa\ra^0,\\
&J_{(2,2)^0}^{(2)}= \left(-i(b+b^{-1}) w_{-2}-w_{-1}^2+h_{-1}^2\right)|\kappa\ra^0.\notag
\end{align*}
The first examples in the space $\calF\otimes\mathrm{F}_{2}$ are:
\begin{align*}
&J_{(1)^0}^{(2)}=D^2|\kappa\ra^0, \quad J_{(3)^0}^{(2)}=-(i w_{-1}-b^{-1}h_{-1})D^2|\kappa\ra^0, \quad J_{(1,1,1)^0}^{(2)}= -(i w_{-1}-bh_{-1})D^2|\kappa\ra^0,\notag\\
&J_{(1,1,1,1,1)^0}^{(2)}=(-2ib w_{-2}-w_{-1}^2-2ib w_{-1}h_{-1}+b h_{-1}^2+2b h_{-2})D^2|\kappa\ra^0, \notag \\
&J_{(2,2,1)^0}^{(2)}=\left(i(b-b^{-1}) w_{-2}-w_{-1}^2-2ib w_{-1}h_{-1}+b^2h_{-1}^2+(1-b^2) h_{-2}\right)D^2|\kappa\ra^0,\\
&J_{(3,1,1)^0}^{(2)}= \left(-i(b+b^{-1}) w_{-1}h_{-1}-w_{-1}^2+h_{-1}^2\right)D^2|\kappa\ra^0.\notag
\end{align*}
Note that in the examples above the coefficients of $J^{(2)}_\lambda$ in terms of $h_n, w_n$ become  simpler.

%It remains to give some formulae for $J^{(2)}_{\widetilde{\lambda}}|\kappa\ra^{\sigma}$, where $\widetilde{\lambda}$ is a 2-core. By calculation in $q=0$ module
%\begin{align}
%& e_{1} J^{(2)}_{(2,1)}=J^{(2)}_{\varnothing}, \qquad
%e_{-1} J^{(2)}_{\varnothing}=J^{(2)}_{(1)}, \qquad
%e_{1} J^{(2)}_{(1)}=J^{(2)}_{(3,2,1)}, \\
%& f_{-1} J^{(2)}_{(\varnothing)}=J^{(2)}_{2,1}, \qquad
%f_{1} J^{(2)}_{1}=J^{(2)}_{(\varnothing)}, \qquad
%f_{3} J^{(2)}_{(3,2,1)}=J^{(2)}_{(1)},
%\end{align}
%Hence we assume an explicit formulae:
%\begin{align}
%J^{(2)}_{(2n+1,2n,\dots,1)}|\kappa\ra^0=e_{2n+1}e_{2n-1}\cdots e_1 |\kappa\ra^0, \qquad
%J^{(2)}_{(2n,2n-2,\dots,1)}|\kappa\ra^0=f_{2n-1}\cdots f_1 |\kappa\ra^0
%\end{align}

\subsection{Vertex operator}

In this subsection we study the vertex operators for the deformed algebra $\calE_1(q,t)$ and then pass to the limit $q,t \rightarrow -1$. Remarkably this limit operator is a natural vertex operator for the algebra $\calA(1,2)=\calH\oplus \Asl(2)_1$.

\secpart Awata et al \cite{AFHKSY:2011} introduced the vertex operators for the algebra $\calE_1(q,t)$. As usual these operators can be defined by commutation relations with the generators of the algebra. The level~1 vertex operator $\Phi(z)\colon \calF_u \rightarrow \calF_v$ defining relations read:
\begin{align}
&(1-v\frac{w}z)E(z)\Phi(w)=(1-q^{-1}tv\frac{w}z)\Phi(w)E(z),\notag \\
&(1-(t/q)^{3/2}u\frac{w}z)F(z)\Phi(w)=(1-(t/q)^{1/2}uw/z)\Phi(w)F(z),
\notag \\
&(1-(t/q)^{-1/4}vw/z)(1-(t/q)^{7/4}u\frac{w}z)\psi^+(z)\Phi(w)\notag \\
&\qquad=
(1-(t/q)^{3/4}v\frac{w}z)(1-(t/q)^{3/4}u\frac{w}z)\Phi(w)\psi^+(z),
\label{eqDI_Vertex_comm} \\
&(1-(t/q)^{1/4}v\frac{w}z)(1-(t/q)^{5/4}u\frac{w}z)\psi^-(z)\Phi(w) \notag \\
&\qquad=
(1-(t/q)^{5/4}v\frac{w}z)(1-(t/q)^{1/4}u\frac{w}z)\Phi(w)\psi^-(z). \notag
\end{align}
In the level 1 case authors of \cite{AFHKSY:2011} provided the explicit exponential form for this operator:
\begin{align}\label{eqVertex_DI}
\Phi(z)= \exp\left(-\sum_{n=1}^{\infty}
\frac{v^{n}-(t/q)^{n}u^{n}}{1-q^{n}}\cdot
\frac{a_{-n}z^{n}}{n}\right)\exp \left(\sum_{n=1}^{\infty}
\frac{v^{-n}-u^{-n}}{1-q^{-n}} \cdot \frac{a_{n}z^{-n}}{n}\right).
\end{align}
The following proposition was stated in \cite{AFHKSY:2011}.

\begin{Proposition} The matrix elements of the operator $\Phi(z)$ in basis $J_\lambda$ have factorized (Nekrasov) form
\begin{align}\label{eqMatrElem_DI}
\langle J_{\lambda} |\Phi(z) |J_{\mu} \rangle =
N_{\lambda,\mu} \left(\frac{qv}{tu}\right)\cdot
\left(\frac{tu}{q}\right)^{|\lambda|}\left(-\frac{v}{q}\right)^{-|\mu|}
t^{n(\lambda)}q^{n(\mu')}z^{|\lambda|-|\mu|},
\end{align}
where $n(\lambda)$ defined in \eqref{eqn_lambda} and
\begin{align}
N_{\lambda,\mu}(u)= \prod_{s \in \lambda}
(1-uq^{-a_{\mu}(s)-1}t^{-l_{\lambda}(s)})\cdot \prod_{t\in \mu}
(1-uq^{a_{\lambda}(t)}t^{l_{\mu}(t)+1}) \label{eqN_lambda_mu}.
\end{align}
arm $a_{\lambda}(s)$ and leg $l_{\lambda}(s)$ lengths defined in \eqref{eqarm_leg}.
\end{Proposition}

\secpart The vertex operator \eqref{eqVertex_DI} depends on parameters $u$ and $v$. We parameterize them by $u=(-1)^{\sigma} e^{\kappa\tau}$ and $v=(-1)^{\tilde{\sigma}} e^{\tilde{\kappa}\tau}$ and then pass to the limit $\tau \rightarrow 0$.  We have two natural cases: in the first case $\sigma=\tilde{\sigma}$, in the second $\sigma\neq \tilde{\sigma}$.
\begin{align}
\Phi_{0}(z) \colon &\calF^{(0)}(\kappa) \rightarrow \calF^{(0)}(\tilde{\kappa}),\; \calF^{(1)}(\kappa) \rightarrow \calF^{(1)}(\tilde{\kappa})  \quad \Phi_{1}(z) \colon \calF^{(0)}(\kappa) \rightarrow \calF^{(1)}(\tilde{\kappa}),\; \calF^{(1)}(\kappa) \rightarrow \calF^{(0)}(\tilde{\kappa})   \notag \\
\Phi_{0}(z)=&\exp\Big(i(\alpha-Q)\sum_{n=1}^{\infty}
\frac{a_{-2n}z^{2n}}{-2n}\Big)\exp\Big(i\alpha\sum_{n=1}^{\infty}
\frac{a_{2n}z^{-2n}}{2n}\Big)=\mathcal{V}_{\alpha}, \label{eqPhi_0} \\
\Phi_{1}(z)=&\exp\Big(i(\alpha-Q)\sum_{n=1}^{\infty}
\frac{a_{-2n}z^{2n}}{-2n}\Big)\exp\Big(i\alpha\sum_{n=1}^{\infty}
\frac{a_{2n}z^{-2n}}{2n}\Big)\cdot\exp\left( \sum_{n} \dfrac{a_{2n+1}}{-2n-1} z^{-2n-1} \right)=\notag \\
=&\mathcal{V}_{\alpha}\cdot\exp(\phi) = \mathcal{V}_{\alpha}\cdot\mathcal{W}, \label{eqPhi_1}
\end{align}
where $\alpha=\dfrac{\kappa-\tilde{\kappa}}{\sqrt{\eps_1\eps_2}}$. Note that $\mathcal{V}_{\alpha}$ is the standard rotated vertex operator for the Heisenberg algebra $\mathcal{H}$ (see \cite{Carlson_Okounkov:2008}). The operator $\mathcal{W}$ is the vertex operator which acts on $\widehat{\mathfrak{sl}}(2)_1$ part. This vertex operator permutes $\calL_{0,1}$ and $\calL_{1,1}$.

Note that formula \eqref{eqPhi_1} is similar to eq. (1) in \cite{Serban:1999}.

\secpart Recall that in the limit $\tau \rightarrow 0$ the basis vectors $J_{\lambda}(q,t)$ tend to 0 as $\tau^{|\lambda^\lozenge|}$. For the limit of the matrix element we can have three possibilities. In the first, the matrix elements $N_{\lambda,\mu}(qv/tu)$ (see \eqref{eqMatrElem_DI}) tend to 0 as $\tau^{|\lambda^\lozenge|+|\mu^\lozenge|}$. In this case the limit of the matrix element is nonzero. If $N_{\lambda,\mu}(qv/tu)$ tends to 0 faster than $\tau^{|\lambda^\lozenge|+|\mu^\lozenge|}$ then the limit of the matrix element is zero. At last, if $N_{\lambda,\mu}(qv/tu)$ tends to 0 slower  than $\tau^{|\lambda^\lozenge|+|\mu^\lozenge|}$ we have a pole in the matrix element.

By combinatorial consideration one can see that the only first two options exist in our case. The precise statements are given in the  propositions below. We will use:
\begin{align} \label{eq:N(2)}
N^{(2)}_{\lambda^{\tilde{\sigma}}, \mu^{\sigma}}(\alpha)=\prod_{S(\lambda^{\tilde{\sigma}}, \mu^{\sigma})}(-b l_{\lambda}(s)+\frac{a_{\mu}(s)+1}{b}-\alpha) \, \cdot \prod_{S(\mu^\sigma,\lambda^{\tilde{\sigma}})}(b(l_{\mu}(t)+1)-\frac{a_{\lambda}(s)}b+\alpha),
\end{align}
where
\begin{align}\label{eq:S(lamba,mu)}
s\in S(\lambda^{\tilde{\sigma}}, \mu^{\sigma})  \Longleftrightarrow s\in \lambda, \; \text{and}\;
l_\lambda(s)+a_\mu(s)+1+\sigma-\tilde{\sigma}\equiv 0 (\mathrm{mod}\, 2).
\end{align}

\begin{Proposition} \label{Prop_ME_0}
 Let $\sigma=\tilde{\sigma}$; $\Phi_{0}(z)\colon \calF^{\sigma}(\kappa_1) \rightarrow \calF^{\tilde{\sigma}}(\kappa_2)$. Then the limit $$\lim\limits_{\tau \rightarrow 0}\left(\tau^{-|\lambda^\lozenge|-|\mu^\lozenge|} N_{\lambda,\mu}(qv/tu)\right) \neq 0\quad \text{iff}\quad d(\lambda^\sigma)=d(\mu^\sigma).$$ In this case matrix element equals:
\begin{align*}
\langle J_{\lambda}^{(2)} |\Phi_{0}(z) |J_{\mu}^{(2)} \rangle=(-1)^{|\mu^\lozenge|} z^{|\lambda^\lozenge|-|\mu^\lozenge|} N^{(2)}_{\lambda^\sigma,\mu^{\tilde{\sigma}}}(\alpha).
\end{align*}
\end{Proposition}

\begin{Proposition} \label{Prop_ME_1}
Let $\sigma-\tilde{\sigma}\equiv 1 \!\!\mod
2$; $\Phi_{1}(z)\colon \calF^{\sigma}(\kappa_1) \rightarrow \calF^{\tilde{\sigma}}(\kappa_2)$. Then the limit $$\lim\limits_{\tau \rightarrow 0}\left(\tau^{-|\lambda^\lozenge|-|\mu^\lozenge|} N_{\lambda,\mu}(qv/tu)\right) \neq 0\quad \text{iff}\quad 2d(\lambda^\sigma)+\tilde{\sigma}=2d(\mu^{\tilde{\sigma}})+\sigma\pm1 .$$ In this case matrix element equals:
\begin{align}\label{eqMatrElem_Phi1}
\langle J_{\lambda}^{(2)} |\Phi_{1}(z) |J_{\mu}^{(2)} \rangle=(-1)^{|\mu-\mu^\lozenge|} z^{|\lambda^\lozenge|-|\mu^\lozenge|} N^{(2)}_{\lambda^\sigma,\mu^{\tilde{\sigma}}}(\alpha).
\end{align}
\end{Proposition}
The combinatorial conditions in the Propositions \ref{Prop_ME_0} and \ref{Prop_ME_1} have a clear algebraic meaning. Namely the condition $d(\lambda^\sigma)=d(\mu^\sigma)$ means that the $h_0$ gradings on $J^{(2)}_{\lambda^\sigma}$ and $J^{(2)}_{\mu^\sigma}$ coincide. The condition $2d(\lambda^{\tilde{\sigma}})+\tilde{\sigma}= 2d(\mu^{\sigma})+\sigma\pm1$ means that the $h_0$ gradings on $J^{(2)}_{\lambda^\sigma}$ and $J^{(2)}_{\mu^\sigma}$ differs on 1. In particular, we see that the operator $\mathcal{W}$ shifts the $h_0$ grading by 1.

\secpart The vertex operators $\Phi_0, \Phi_1$ (\eqref{eqPhi_0}, \eqref{eqPhi_1})  can be also rewritten in terms of $h_n$ and $w_n$. It is evident that $\Phi_0$ can be rewritten:
\begin{align}\label{eqPhi_0_in_h}
\Phi_{0}(z)=\mathcal{V}_{\alpha}(z)=\exp\Big(i(\alpha-Q)\sum_{n=1}^{\infty}
\frac{w_{-n}z^{2n}}{-2n}\Big)\exp\Big(i\alpha\sum_{n=1}^{\infty}
\frac{w_{n}z^{-2n}}{2n}\Big).
\end{align}

In order to rewrite $\Phi_1$ we need to rewrite the $\widehat{\mathfrak{sl}}(2)$ nontrivial part $\mathcal{W}$ \eqref{eqPhi_1}. Introduce two operators:
\begin{align}
\mathcal{W_+}(z)=z^{h_0/4}D\exp\Bigl(\sum_{n \in \mathbb{Z}_{>0}} \dfrac{h_{-n}}{2n} z^{n} \Bigr) \exp\Bigl(\sum_{n \in \mathbb{Z}_{>0}} \dfrac{h_n}{-2n} z^{-n} \Bigr)z^{h_0/4}=:\!\exp\Bigl(\varphi(z)\Bigr)\!:,\notag \\
\mathcal{W_-}(z)=z^{-h_0/4}D^{-1}\exp\Bigl(\sum_{n \in \mathbb{Z}_{>0}} \dfrac{h_{-n}}{-2n} z^{n} \Bigr) \exp\Bigl(\sum_{n \in \mathbb{Z}_{>0}} \dfrac{h_n}{2n} z^{-n} \Bigr)z^{-h_0/4}=:\!\exp\Bigl( -\varphi(z)\Bigr)\!:,   \label{eqW_h_n}
\end{align}
where $\varphi(z)$ defined in \eqref{eq:varphi}. The operator $\mathcal{W}_+$ increases the $h_0$ grading by 1 and the operator $\mathcal{W}_-$ decreases the $h_0$ grading by 1. Under the operator-state correspondence these operators correspond to the vectors $|\kappa\ra^1$ and $f_{-1}|\kappa\ra^1$ (equivalently $J_{\varnothing^1}^{(2)}$ and $J_{(1)^1}^{(2)}$) on the top of the representation $\calL_{1,1}$,  see picture \ref{fig_11}).
\begin{Proposition} \label{Prop:W}
The operator $\mathcal{W}$ has the form
\begin{align} \label{eq:W=W+W-}
\mathcal{W}= z^{1/2}\mathcal{W}_-(z^2)+z^{-1/2}\mathcal{W}_+(z^2)
\end{align}
\end{Proposition}
The rescaling $z \rightarrow z^2$ and the factors $z^{\pm1/2}$ are just the transformation from the homogeneous grading to the principal grading. If we consider $\mathcal{W}\colon \calL_{0,1} \rightarrow \calL_{1,1}$ then $z^{-1/2}\mathcal{W}_+(z^2)$ consists of terms of even degrees of $z$ and $z^{1/2}\mathcal{W}_-(z^2)$ consists of terms of odd degrees of $z$. If we consider $\mathcal{W}\colon \calL_{1,1} \rightarrow \calL_{0,1}$ then $z^{-1/2}\mathcal{W}_+(z^2)$ consists of terms of odd degrees of $z$ and $z^{1/2}\mathcal{W}_-(z^2)$ consist of terms of  even degrees of $z$.

Therefore the operators $\mathcal{V}_{\alpha}(z)\mathcal{W}_+(z)$ and $\mathcal{V}_{\alpha}(z)\mathcal{W}_+(z)$ have factorized matrix elements given by the formula \eqref{eqMatrElem_Phi1} or 0. Recall that the matrix elements of $\mathcal{V}_{\alpha}(z)\mathcal{W}_+(z)$ are nonzero if the $h_0$ grading is increased by 1 and the matrix elements of $\mathcal{V}_{\alpha}(z)\mathcal{W}_-(z)$ are nonzero if the $h_0$ grading is decreased  by 1.

\section{$r=2$ case}
\subsection{Algebras, representations, characters}

\secpart As was already mentioned in the Introduction we want to construct the special (geometric) basis in the representations of the algebra $\calA(2,2)=\calH\oplus \Asl(2)_2 \oplus \NSR$. In this paper we consider only the Neveu--–Schwarz sector of this algebra i.e. the $\NSR$ algebra is generated by $L_n, G_r$, $n \in \Z, r \in \Z+\frac12$ subject of relations:
\begin{align}
&[L_{n},L_{m}]=(n-m)L_{n+m}+\frac{c_{\textsf{\tiny{NSR}}}}{8}(n^{3}-n)\delta_{n+m},\notag \\
&\{G_{r},G_{s}\}= 2L_{r+s}+\frac{1}{2}c_{\textsf{\tiny{NSR}}}(r^{2}-\frac{1}{4})\delta_{r+s}, \label{eq:NSR}\\
&[L_{n},G_{r}]=\left(\frac{1}{2}n-r\right)G_{n+r}, \notag
\end{align}
where the central charge $c_{\scriptscriptstyle{\textsf{NSR}}}$ is parameterized as follows
\begin{align*}
c_{\textsf{\tiny{NSR}}}=1+2Q^{2}, \qquad Q=b+\frac{1}{b}.
\end{align*}

We will consider the representation of the algebra $\calA=\calH\oplus \Asl(2)_2 \oplus \NSR$ which is a tensor product of representations of the algebras $\calH$, $\Asl(2)_2$ and $\NSR$. For the Heisenberg algebra $\calH$ we take standard Fock module $\calF$ with the character $\chi_B$, see \eqref{eqchi_Asl1}. For the $\NSR$ algebra we take the Verma module with the highest weight parameterized by the parameter~$P$, namely the highest weight vector is denoted by $|P\ra$ and defined by properties
\begin{align*}
    L_{n}|P\rangle=G_{r}|P\rangle=0\quad\text{for}\quad n,r>0,\qquad
    L_{0}|P\rangle=\Delta_{\textsf{\tiny{NS}}}(P)|P\rangle,
\end{align*}
where $\Delta_{\textsf{\tiny{NS}}}(P)=\frac{1}{2}(Q^2/4-P^2)$. We denote this representation as $\sfV_{\Delta_{\textsf{\tiny{NS}}}}$ or simply  $\sfV_\Delta$. The character of $\sfV_\Delta$ equals \begin{align}\label{eqchi_NSR}
\chi_{\NSR}(q)=\left.\text{Tr}q^{L_0}\right|_{\sfV_\Delta} =q^{\Delta}\chi_B(q) \chi_F(q),\quad \text{where}\quad  \chi_F(q)=\!\!\!\!\!\!\prod_{r+\frac12 \in \Z_{>0}} \!(1+q^r) .
\end{align}

The $\Asl(2)$ algebra has three integrable representations of level 2: $\calL_{0,2}$, $\calL_{1,2}$ and $\calL_{2,2}$. The representation $\calL_{0,2}\oplus\calL_{2,2}$ is called the Neveu--–Schwarz representation of $\Asl(2)_2$. The character of this representation equals \cite{GKO}[eq 4.2]
\begin{align}\label{eqchi_Asl2}
\!\!\!\!\chi_{\Asl}^{(0,2)}(q,t)+\chi_{\Asl}^{(2,2)}(q,t)= \!\!\!\!\!\! \prod_{r+\frac12 \in \Z_{>0}} \!\!\!\!\!\!(1+t^{-1}q^r)(1+q^r)(1+tq^r) \!=\! 	 \sum_{n \in \Z} t^nq^{n^2/2}\chi_B(q) \chi_F(q).
\end{align}

Note that the remaining representation $\calL_{1,2}$ is called the Ramond representation of~$\Asl(2)_2$.

\secpart \label{sub:r=2:char} We want to construct a basis in the representation
\begin{align}\label{eq:r2:repr}
\calF \otimes \left( \calL_{0,2} \oplus \calL_{2,2} \right) \otimes \sfV_\Delta
\end{align}
of the algebra $\calA(2,2)$ with the properties from the Introduction. The basic vectors are labeled by torus fixed points on the moduli space $\bigsqcup_N \calM(2,N)^{\Z_2}$. In this case the torus fixed points are labeled by pairs of Young diagrams $\lambda_1,\lambda_2$  colored in two colors with the angles colored in the same color $\sigma$. We will denote the corresponding basic vector by $J_{\lambda_1^\sigma,\lambda_2^\sigma}$ or $J_{\vec{\lambda}^\sigma}$ for short. As in $r=1$ case such a basis respects grading:
\begin{align} \label{eq:r2:h0}
h_0(J_{\vec{\lambda}^\sigma})=(2d(\vec{\lambda}^\sigma)+ 2\sigma)J_{\vec{\lambda}^\sigma}, \qquad L_0(J_{\vec{\lambda}^\sigma})=(\frac{2|\vec{\lambda}^\sigma|+h_0}4+\Delta)J_{\vec{\lambda}^\sigma},
\end{align}
where $d(\vec{\lambda}^\sigma)=d(\lambda_1^\sigma)+ d(\lambda_2^\sigma).$

The existence of this basis means that we can compute the character of representation of the algebra $\calA(2,2)$. Using the formula \eqref{eqchi^(1)} we get:
\begin{align*}
&\chi^{(2)}(q,t):=\sum_{\sigma=0,1} \sum_{\lambda_1^\sigma,\lambda_2^\sigma} q^{\frac{|\vec{\lambda}|+d(\vec{\lambda})+\sigma}2+\Delta}t^{d(\vec{\lambda})+\sigma}= q^\Delta \!\!\sum_{\sigma, d_1, d_2} (t\sqrt{q})^{d_1+d_2+\sigma} \chi^{(1)}_{\sigma,d_1}(q)\chi^{(1)}_{\sigma,d_2}(q) = \notag \\
=&q^{\Delta}\chi_B(q)^4\sum_{d_1,d_2} \left( (t\sqrt{q})^{d_1+d_2} q^{\frac{2d_1^2+2d_2^2-d_1-d_2}2}+(t\sqrt{q})^{d_1+d_2+1} q^{\frac{2d_1^2+2d_2^2+d_1+d_2}2}\right)= \\ =&q^{\Delta}\chi_B(q)^4\sum_d (t\sqrt{q})^{d} \sum_{d_1} \left( q^{\frac{(2d_1-d)^2+d^2-d}2}+q^{\frac{(2d_1-d+1)^2+d^2-d}2} \right)= q^{\Delta}\chi_B(q)^4\sum_{d,k}q^{\frac{d^2+k^2}2}t^d. \notag
\end{align*}
Using the Jacobi triple product identity this expression can be rewritten as
\begin{align*}
\chi^{(2)}(q,t)=&q^{\Delta}\chi_B(q)^4\sum_{d,k}q^{\frac{d^2+k^2}2}t^d= q^{\Delta}\chi_B(q)^2 \prod_{r+\frac12 \in \Z_{>0}}(1+t^{-1}q^r)^2(1+q^r)(1+tq^r)=\notag \\
=&\chi_B(q) \cdot\left(\chi_{\Asl}^{(0,2)}(q,t)+\chi_{\Asl}^{(2,2)}(q,t)\right)\cdot q^{\Delta}\chi_F(q)\chi_B(q)=
\chi_{\calH\oplus \Asl(2)_2 \oplus \NSR},
\end{align*}
where we used the formulas \eqref{eqchi_NSR} and \eqref{eqchi_Asl2}.

\secpart The generators of $\mathcal{H}$ are denoted by $w_n$ as in $r=1$ case but we rescale commutation relations to:
\begin{align*}
[w_n,w_m]=4n\delta_{n+m}.
\end{align*}
So the full set of the generators of the algebra $\calA(2,2)$ consists of $w_n; e_n, f_n, h_n; L_n, G_r$.

We will use the free field realization of the $\NSR$ and $\Asl(2)_2$ algebras. The free field realization means that we consider the Fock representation of the Heisenberg algebra and the Majorana fermion algebra and define the action of the mentioned before algebras on this representation. This method is very useful since the Heisenberg and the  Majorana fermion algebras are much simpler.

Recall the Fateev--Zamolodchikov \cite{Fateev:1985mm} realization of the algebra $\widehat{\mathfrak{sl}}(2)_2$. Introduce generators $h_n, D, \chi_r$:
\begin{align*}
[h_{n},h_{m}]= 4n\delta_{n+m,0}, \quad
\{\chi_{r},\chi_{s}\} =\delta_{r+s,0}, \quad [h_n,\chi_r]=0.
\end{align*}
Clearly $h_n$ generate the Heisenberg algebra $\calH$, and $\chi_r$ generate the Fermion algebra $\mathsf{F}$. Denote by $\rmF_{P}$ the Fock representation of algebra $\calH\oplus \mathsf{F}$ with vacuum vector $v_P$:
\begin{equation*}
h_nv_P=0,\,\chi_rv_p=0\quad \text{for $n,r>0$};\qquad h_0v_P=Pv_P.
\end{equation*}
As in $r=1$ case we introduce the operator $D\colon \rmF_{P} \rightarrow \rmF_{P+1}$  defined by commutation relations
\begin{equation*}
[D,h_n]=0\; \text{for}\; n\ne 0, \quad [h_0,D]=D,\quad [\chi_r,D]=0.
\end{equation*}
Then $\widehat{\mathfrak{sl}}(2)_2$ representations $\mathcal{L}_{0,2} \oplus \mathcal{L}_{2,2}$ can be realized as the direct sum:
\begin{align}
\mathcal{L}_{0,2} \oplus \mathcal{L}_{2,2}=\bigoplus_{d \in \mathbb{Z}} \rmF_{2d}. \label{eq:L02+L22}
\end{align}
The operators $e_n, f_n$ are defined by
\begin{align*}
e(z)=\sum_{n \in \mathbb{Z}} e_nz^{-n}=\chi(z):\!\exp\left(2 \varphi(z)\right)\!:,\quad
f(z)=\sum_{n \in \mathbb{Z}} f_nz^{-n}=\chi(z):\!\exp\left(-2 \varphi(z)\right)\!:,
\end{align*}
where $\chi(z)=\sum_r \chi_r z^{-r-1/2}$ and $\varphi(z)$ is defined similar to \eqref{eq:varphi} by the formula:
\begin{align}
\varphi(z)=\sum_{n \in \mathbb{Z}\setminus 0} \dfrac{h_n}{-4n} z^{-n}+\frac{h_0\log z}4+\widehat{Q},
\end{align}

For the $\NSR$ algebra we consider the Fock representation of the algebra with free boson generators $c_n$, $n \in \Z\setminus\{0\}$ and free fermion generators $\psi_r$, $r \in \Z+\frac12$
\begin{align}
&[c_{n},c_{m}]= n\delta_{n+m,0}, \qquad
&\{\psi_{r},\psi_{s}\} =\delta_{r+s,0}
\end{align}
The $\NSR$ algebra is embedded into the (completed) universal enveloping algebra of $\la c_m,\psi_r \ra$ by formulas:
\begin{equation} \label{eq:NSR:bos}
\begin{aligned}
&L_{n}= \frac{1}{2}\sum_{k\neq
0,n}c_{k}c_{n-k}+\frac{1}{2}\sum_{r}(r-\frac{n}{2})\psi_{n-r}\psi_{r}
+\frac{i}{2}(Qn\pm 2P)c_{n} \\
&L_{0}=\sum_{k>0}c_{-k}c_{k}+\sum_{r>0}r\psi_{-r}\psi_{r}+\frac{1}{2}\big(\frac{Q^{2}}{4}-P^{2}\big)  \\
&G_{r}= \sum_{n\neq 0}c_{n}\psi_{r-n}+i(Qr\pm P)\psi_{r}.
\end{aligned}
\end{equation}
Therefore $\NSR$ acts on the Fock representation of the algebra generated by $c_n,\psi_r$ and the corresponding representation is isomorphic to a Verma module $\sfV_{\Delta}$ for general values of $b,P$.

After all, in the free realization of the representation \eqref{eq:r2:repr} of the algebra $\calA(2,2)$ we have generators $w_n; h_n, D, \chi_r; c_n, \psi_r$. It follows from \eqref{eq:L02+L22} that this representation has a decomposition:
\begin{align} \label{eq:r2:decomposition}
\calF \otimes \left( \calL_{0,2} \oplus \calL_{2,2} \right) \otimes \sfV_\Delta=\calF\otimes \bigoplus_{d \in \mathbb{Z}} \rmF_{2d} \otimes \sfV_\Delta.
\end{align}
Each summand on the right hand side is a representation of the subalgebra $\calH\oplus \calH \oplus \mathsf{F}\oplus \NSR$.
The highest weight vectors for this subalgebra can be obtained by action of $D$ to the vacuum. And we want to express the basis $J_{\vec{\lambda}^\sigma}$ in these generators.

\subsection{The limit of the algebra}
Our strategy will be the same as in the previous section: we will study the limit of $\calE_1(q,t)$. In this case we will need a level two representations of this algebra. We argue that in the limit we will get the $\calA(2,2)$ algebra.

\secpart The algebra $\calE_1$ has a formal Hopf algebra structure.
The formulas for the coproduct read $\Delta(C)=C  \otimes C $ and
\begin{equation} \label{eqCopro}
\begin{aligned}
&
\Delta (\psi^\pm(z))=
 \psi^\pm \left(C_{(2)}^{\pm 1}\cdot z)\right)\cdot \psi^\pm \left(C_{(1)}^{\mp 1}\cdot z\right), \\
&\Delta (E(z))= E(z)\otimes 1+
  \psi^-\left(C_{(1)} \cdot z\right)\cdot E\left( C_{(1)}^2\cdot z\right), \\
&\Delta (F(z))=
  F(C_{(2)}^2z)\otimes \psi^+(C_{(2)}z)+1 \otimes F(z),
\end{aligned}
\end{equation}
where $C_{(1)} \seteq C\otimes 1$
and   $C_{(2)} \seteq 1\otimes C$.
Since we do not use the antipode and the counit in this paper, we omit them.

Using the coproduct map $\Delta$ one can define the representation of $\calE_1$ in the space
\begin{align} \label{eq:Fu1u2}
\calF_{\vec{u}}=\calF_{u_1}\otimes \calF_{u_2}.
\end{align}
Evidently this is a representation of the level $2$. For the bosonization of this algebra we will need two Heisenberg algebras: say $a_n^{(1)}$ for $\calF_{u_1}$ and $a_n^{(2)}$ for $\calF_{u_2}$. For example using the coproduct formulas \eqref{eqCopro} and realization formulas \eqref{eqDI_Fock} one can realize $E(z)$ in terms of $a_n^{(1)}$ and $a_n^{(2)}$:
\begin{equation*}
\begin{aligned}
E(z)=&u_1:
\exp\Big(-\sum_{n\in \Z} \dfrac{1-t^{n} }{n}a^{(1)}_n z^{-n} \Big):+\\
+&u_2\exp\Big(
 \sum_{n=1}^{\infty} \dfrac{1-t^{-n}}{n} (1-t^n q^{-n}) a_{-n}^{(1)}z^{n}
    \Big)
:\!\!\exp\Big(-\sum_{n \in \Z} \dfrac{1-t^{n} }{n}(t/q)^{-n/2} a^{(2)}_n z^{-n} \Big)\!\!:
\end{aligned}
\end{equation*}

\secpart \label{sub:r=2:tens} Now we consider the limit:
\begin{align*}
q=-e^{-\tau \eps_{2}},\; t=-e^{\tau \eps_{1}},\; u_{i} = (-1)^{\sigma}e^{\tau \kappa_{i}} \quad \tau \rightarrow 0.
\end{align*}
Denote by $\calF^{(\sigma,\sigma)}(\kappa_1,\kappa_2)$ \label{sub:calF2} the limit of the representation  $\calF_{u_1}\otimes \calF_{u_2}$. We will consider the direct sum
\begin{align*}
\calF^{(0,0)}(\kappa_1,\kappa_2)\oplus\calF^{(1,1)}(\kappa_1,\kappa_2).
\end{align*}
This space will have a basis labeled  by pairs of two colors colored Young diagrams $\lambda_1,\lambda_2$ with angles colored in the same color $\sigma$.  It was shown in subsubsection \ref{sub:r=2:char} that the character of this representation coincides with the character of \eqref{eq:r2:decomposition}.

Similar to the $r=1$ case (formulas \eqref{eqr=1_limit}) one can find in the space $\calF^{(\sigma,\sigma)}(\kappa_1,\kappa_2)$
\begin{align}
&E(z), F(z) \mapsto
:(-1)^\sigma\left(\exp(2\phi^{(1)}(z))+\exp(2\phi^{(2)}(z))\right) :+O(\tau), \notag  \\
&\frac{\psi^\pm(z)-\psi^\pm(-z)}2 \mapsto \mp \left(2Q\sum_{\pm n \in \Z_{>0}} a_{2n+1}    z^{-(2n+1)}\right) \tau +O(\tau^2), \label{eqr=2_limit}\\
&\frac{\psi^\pm(z)+\psi^\pm(-z)}2 \mapsto 1+ \left(2Q^2(\sum_{\pm n \in \Z_{>0}} a_{2n+1}    z^{-(2n+1)})^2-2Q\epsilon_1 \sum_{\pm n \in \Z_{>0}} 2n a_{2n} z^{-2n}\right)\tau^2+O(\tau^3), \notag
\end{align}
where
\begin{align} \label{eq:phi^(k)}
a_{2n+1}=a_{2n+1}^{(1)}+a_{2n+1}^{(2)},\quad\phi^{(k)}(z)=\sum_{n} \dfrac{a_{2n+1}^{(k)}}{-2n-1} z^{-2n-1},\;\; k=1,2.
\end{align}
The operators $a_{2n+1}$ can be considered as the action of generators of $\Asl(2)_2$ (see \eqref{eqGenLW}) on the tensor product $\calL_{\sigma,1}\otimes \calL_{\sigma,1}$. Then the limit of $E(z)$ coincides with the $b(z)=2\sum_n b_nz^{-n}$, where operators $b_n$ are the action of generators of $\Asl(2)_2$ (see \eqref{eqGenLW}) on the same tensor product. The operators $a_{2n}$ can be identified with the Heisenberg generators $w_n$. In other words we have found the algebra $\calH \oplus \Asl(2)_2$ in the limit.

The limit of the algebra $\calE_1(q,t)$ should be greater then $\calH \oplus \Asl(2)_2$. As was already mentioned in the introduction  we conjecture that in the limit we will see the algebra $\calA(2,2)=\calH\oplus\Asl(2)_2\oplus \NSR$.

In order to support this conjecture one can find operators acting on the representation $\calF^{(\sigma)}(\kappa_1)\otimes \calF^{(\sigma)}(\kappa_1)$ which commutes with $\calH \oplus \Asl(2)_2$. From the conjecture we expect to get one Heisenberg and one fermion algebra.

Indeed the Heisenberg algebra can be given as the simple difference $c_n=\frac12(a_{2n}^{(1)}-a_{2n}^{(2)})$. Evidently these operators commute with
$a_n$ and $b_n$ which generates $\calH \oplus \Asl(2)_2$.

From the other side note that operators $a_{2n+1}^{(1)}, a_{2n+1}^{(2)}$ generate the action of $\Asl(2)_1\oplus\Asl(2)_1$. Then the coset algebra $\dfrac{\Asl(2)_1\oplus\Asl(2)_1}{\Asl(2)_2}$ acts on $\calL_{\sigma,1}\otimes \calL_{\sigma,1}$ and commutes with $\Asl(2)_2$. This coset algebra is known to be a unitary minimal model $M(3,4)$ (\cite{GKO}) and then has the symmetry of the fermion algebra.

\secpart In order to recapitulate the results of the previous subsubsection we present a formulas for the bosonization $w_n; h_n, \chi_r; c_n, \psi_r$, in terms of two limit Heisenberg algebras $a_n^{(1)}$ and $a_n^{(2)}$.
\begin{align}\label{eq:r2:wn}
& w_n=a_{2n}^{(1)}+a_{2n}^{(2)}, \qquad c_n=\frac{a_{2n}^{(1)}-a_{2n}^{(2)}}2 \\&
\sum_n (h_n{-}\delta_{n,0})z^{-2n}=\!\frac{(-1)^{\sigma{+}1}}4\left(
\exp(2\phi^{(1)})+\exp(2\phi^{(2)})+\exp({-}2\phi^{(1)})+\exp({-}2\phi^{(2)} ) \right)
\label{eq:r2:hn} \\&
\sum_r \chi_rz^{-2r}=(-1)^{\sigma+1}\frac{i}{2\sqrt{2}}\left(\exp(\phi^{(1)}+\phi^{(2)})-\exp(-\phi^{(1)}-\phi^{(2)}) \right),
\label{eq:r2:chir} \\&
\sum_r \psi_rz^{-2r}=\frac{i}{2\sqrt{2}}\left(\exp(\phi^{(1)}-\phi^{(2)})-\exp(-\phi^{(1)}+\phi^{(2)}) \right), \label{eq:r2:psir}
\end{align}
where fields $\phi^{(k)}$, $k=1,2$ were introduced in \eqref{eq:phi^(k)}. The explanation of the formulae for fermions $\chi_r,\psi_r$ is given in Appendix~\ref{AppOpeSta}. Anyway it is easy to check that commutation relations of r. h. s. are correct.

\subsection{Basis and the vertex operator for the algebra $\calE_1(q,t)$}

\secpart \label{sub:Plambda} The commutative subalgebra of the algebra $\calE_1(q,t)$ acts on the space $\calF_{u_1}\otimes \calF_{u_2}$. There exists a certain basis $P_{\vec{\lambda}}$ which consists of eigenvectors of the operator $E_0$: \begin{align} \label{eq:r2:E0}
E_0 | P_{\vec{\lambda}}\rangle
=\varepsilon_{\vec{\lambda},\vec{u}}| P_{\vec{\lambda}}\rangle,
 \qquad
 \varepsilon_{\vec{\lambda},\vec{u}}\seteq
 u_1\varepsilon_{\lambda^{(1)}}+u_2\varepsilon_{\lambda^{(2)}}.
\end{align}
 Here $\vec{\lambda}=(\lambda_1,\lambda_2)$ is a pair of partitions. Similar to the Macdonald polynomials this basis can be defined as orthogonalizaiton of a certain monomial basis.
The first examples of the vectors $P_{\vec{\lambda}}$ have the form (\cite{AFHKSY:2011}):
\begin{align}
 &P_{((1),\varnothing)}=P_{(1)}\otimes 1,\qquad
 P_{(\varnothing,(1))}
=1\otimes P_{(1)}
  +(q/t)^{1/2}\dfrac{(t-q)u_2}{q(u_1-u_2)}P_{(1)}\otimes 1 \notag \\
 &P_{((1,1,1),\varnothing)} =P_{(1,1,1)}\otimes 1, \qquad  P_{((2,1),\varnothing)}
=P_{(2,1)}\otimes 1, \qquad P_{((3),\varnothing)} =P_{(3)}\otimes 1, \label{eq:r2:P}
\\
 &P_{((1,1),(1))}
=P_{(1,1)}\otimes P_{(1)}
  +(q/t)^{1/2}\dfrac{(t{-}q)u_2}{q(q u_1{-}u_2)} % \times \sqrt{q/t]
 P_{(2,1)}\otimes 1
+(q/t)^{1/2}\dfrac{(1{-}q)(t{-}q)(1{-}t^3)t^2u_2}{q(1{-}q t^2)(1{-}t)(u_1{-}t^2 u_2)}
P_{(1,1,1)}\otimes 1, \notag
\\
 &P_{((2),(1))}
=P_{(2)}\otimes P_{(1)}
  +(q/t)^{1/2}\dfrac{(t{-}q)u_2}{q(q^2 u_1{-}u_2)} % \times \sqrt{q/t]
 P_{(3)}\otimes 1
+(q/t)^{1/2}\dfrac{(1{-}q^2)(t{-}q)(1{-}q t^2)t u_2}{q(1{-}q t)(1{-}q^2 t)(u_1{-}t u_2)}
   P_{(2,1)}\otimes 1,\notag
\end{align}
where $P_\lambda$ denotes the Macdonald polynomials \eqref{eqMacdonald_in_m}.

It follows from the definition of $P_{\vec{\lambda}}$ that the basis $\{P_{\vec{\lambda}}\}$ is orthogonal. It is clear from the coproduct formula \eqref{eqCopro} that all vectors $P_{\lambda_1,\varnothing}$ have the form $P_{\lambda_1}\otimes 1$.

In was conjectured in \cite{AFHKSY:2011} that there exists another normalization $K_{\vec{\lambda}}$ which is more similar to the integral form of the Macdonald polynomials $J_\lambda(q,t)$. The norms in this  normalization have a nice factorized form:

\begin{align}\label{eq:conj:norm:level:m}
\langle K_{\vec{\lambda}}|K_{\vec{\lambda}}\rangle
=
&\bigl((-1)^m (t/q)^{m-1}u  \bigr)^{|\Vec{\lambda}|}
\times
\prod_{k=1}^m u_k^{-(m-2)|\lambda^{(k)}|}
q^{-(m-2)n(\lambda^{(k)'})}t^{(m-2)n(\lambda^{(k)})}\times
\\
&\prod_{i,j=1}^m
N_{\lambda^{(i)},\lambda^{(j)}}
(q u_i/tu_j), \notag
\end{align}
where $u=u_1u_2$ and $N_{\lambda,\mu}$ is defined in \eqref{eqN_lambda_mu}.

\secpart Vertex operator $\Phi(z)\colon \calF_{\vec{u}} \rightarrow \calF_{\vec{v}}$ is defined by the commutation relations \eqref{eqDI_Vertex_comm}, where $u=u_1u_2$ and $v=v_1v_2$.

 \begin{Conj}[\cite{AFHKSY:2011}] (1) The operator $\Phi(z)$ exists uniquely.

\noindent (2)The matrix elements of the operator $\Phi(z)$ in the basis $K_{\vec{\lambda}}$ have factorized (Nekrasov) form
\begin{align}\label{eq:DI:matrix}
\langle  K_{\vec{\lambda}} |\Phi(z) |K_{\vec{\mu}} \rangle =
&\left(\frac{t^2}{q^2}uv\right)^{|\vec{\lambda}|}\left(\frac{t}{q}v\right)^{-|\vec{\mu}|} z^{|\vec{\lambda}|-|\vec{\mu}|}\times \\
&\prod_{k=1}^2 v_k^{-|\lambda_k|} u_k^{|\mu_k|} q^{-n(\lambda_k'+n(\mu_k'))} q^{n(\lambda_k-n(\mu_k))}   \prod_{i,j=1}^2
  N_{\lambda^{(i)},\mu^{(j)}}\left(\frac{qv_i}{tu_j}\right),\notag
\end{align}
where $u=u_1u_2$, $v=v_1v_2$, $N_{\lambda,\mu}$ is defined in \eqref{eqN_lambda_mu} and $n(\lambda)$ is  defined in \eqref{eqn_lambda}.
\end{Conj}

One can see that in the limit $q,t \rightarrow -1$ of \eqref{eq:DI:matrix} we get the desired expression \eqref{eq:matrix:color} (up to some sign and factors like $2^{|\lambda^\lozenge|}$). So it is natural to consider the limit of the basis $K_{\vec{\lambda}}$.

We remark  that in the case of the level one representations the basis $K_\lambda$ differs from the basis $J_\lambda$ by scalar factor. But this factor is a some power of parameters $t,q,u$ so it is not very important for our limit.

\subsection{Basis for the algebra $\calA(2,2)$}

\secpart We will study the basis in the representation
\begin{align}
\calF^{(0,0)}(\kappa_1,\kappa_2)\oplus\calF^{(1,1)}(\kappa_1,\kappa_2)=\calF \otimes \left( \calL_{0,2} \oplus \calL_{2,2} \right) \otimes \sfV_\Delta \label{eq:r2:rep}
\end{align}
of the algebra $\calA(2,2)$. This representation is called the  Neveu--–Schwarz representation of the algebra $\calA(2,2)$. The equality \eqref{eq:r2:rep} can be considered as the isomorphism between two realizations of the same representation. On the left hand side the algebra $\calA(2,2)$ arises as the limit of $\calE_1(q,t)$, see subsubsection \ref{sub:r=2:tens}. On the right hand side we use the realization $\calH\oplus \Asl(2)_2 \oplus \NSR$.

The continuous parameters $\kappa_1,\kappa_2$ on the left hand side of \eqref{eq:r2:rep} and $\Delta_{\textsf{\tiny{NS}}}$ on the right hand side are related in the realization of $\NSR$ in the limit of $\calE_1(q,t)$. We suggest the relation $P=\frac12(\kappa_2-\kappa_1)$, where $\Delta_{\textsf{\tiny{NS}}}=\frac{1}{2}(Q^2/4-P^2)$. We denote by $|P\rangle$ the vacuum vector of $\calF \otimes \left( \calL_{0,2} \oplus \calL_{2,2} \right) \otimes \sfV_\Delta$. This vector coincides  the vacuum vector with $|\kappa_1,\kappa_2\rangle^{0,0}$ of $\calF^{(0,0)}(\kappa_1,\kappa_2)$. The vacuum vector of $\calF^{(1,1)}(\kappa_1,\kappa_2)$ can be obtained by action of shift operator $D$: $|\kappa_1,\kappa_2\rangle^{1,1}=D|P\rangle$.

By the limit procedure the space \eqref{eq:r2:rep} has a basis $J^{(2)}_{\vec{\lambda}^\sigma}$, labeled by $\sigma=0,1$ and $\vec{\lambda}=(\lambda_1,\lambda_2)$. The desired basis $J^{(2)}_{\vec{\lambda}^\sigma}$ respects the grading by $h_0$ and $L_0$ as in \eqref{eq:r2:decomposition}. For example the action of $h_0$ can be deduced as a limit of the action $E_0$ \eqref{eq:r2:E0} as in $r=1$ case see \eqref{eq:r1:h0}. From the decomposition \eqref{eq:r2:decomposition} follows that the space with the given $h_0$ forms a representation of algebra $\calH\oplus \calH \oplus \mathsf{F} \oplus \NSR$. All the highest vectors for the decomposition \eqref{eq:r2:decomposition} have the form $D^k|P\rangle$, e.g:
\begin{align}
J^{(2)}_{((1)^1,(1)^1)}=D^{-1}|P\rangle, \quad J^{(2)}_{(\varnothing^0,\varnothing^0)}=|P\rangle, \quad J^{(2)}_{(\varnothing^1,\varnothing^1)}=D|P\rangle, \quad J^{(2)}_{((1)^0,(1)^0)}=D^2|P\rangle.
\end{align}

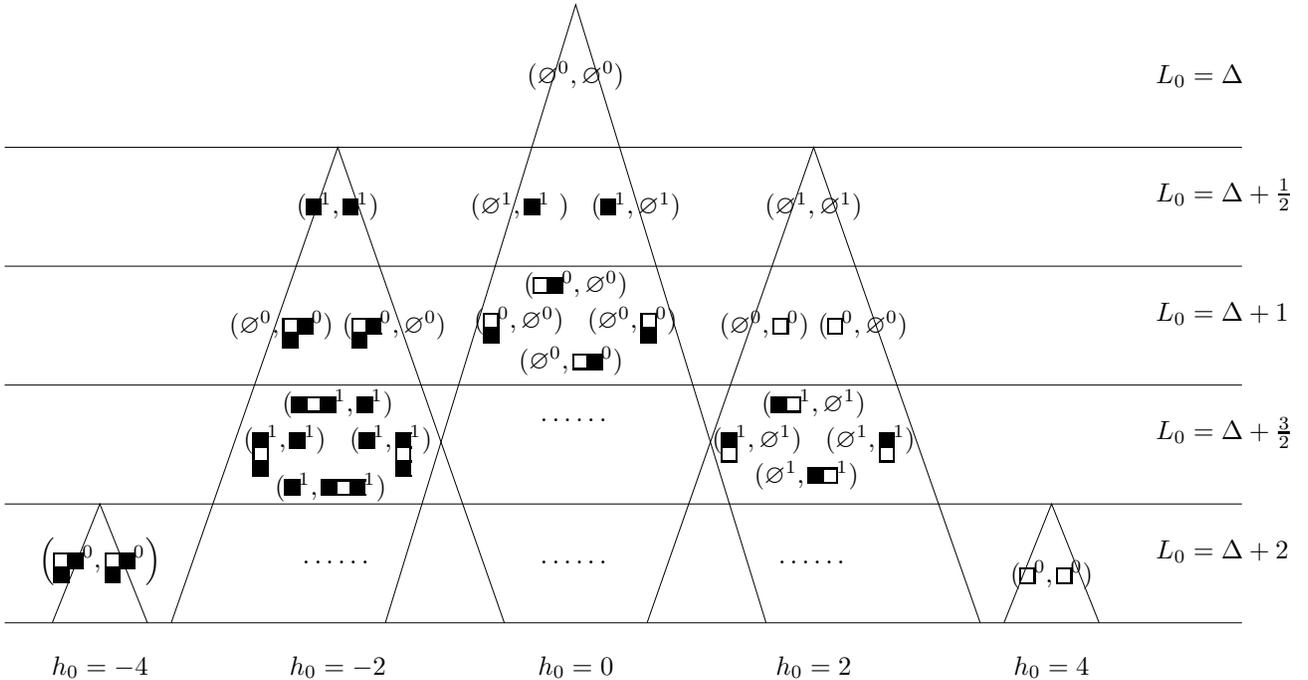
\begin{figure}[h]
\begin{center}
\begin{tikzpicture}[x=1em, y=1em,scale=0.9]
%\ytableausetup{mathmode, boxsize=0.5em}
\draw (0,-1) node[anchor=north]
	{ $(\varnothing^0,\varnothing^0)$};
\draw (10,-6.5) node[anchor=north]
	{$(\varnothing^1,\varnothing^1)$};
\draw (0,-6.5) node[anchor=north]
	{ ($\varnothing^1,
	\begin{ytableau} *(black)\end{ytableau}^1$ )  \,
	$(\begin{ytableau} *(black) \end{ytableau}^1,
	\varnothing^1)$};
\draw (-10,-6.5) node[anchor=north]
	{$(\begin{ytableau} *(black) \end{ytableau}^1,
	\begin{ytableau} *(black)\end{ytableau}^1)$};
\draw (10,-11.5) node[anchor=north]
	{ $(\varnothing^0,
	\begin{ytableau} *(white)\end{ytableau}^0)$
	$(\begin{ytableau} *(white) \end{ytableau}^0,
	\varnothing^0)$};
\draw (0,-9.8) node[anchor=north]
	{$(	\begin{ytableau} *(white) & *(black)\end{ytableau}^0,\varnothing^0)$};
\draw (0,-11.3) node[anchor=north]
	{$(	\begin{ytableau} *(white) \\ *(black)\end{ytableau}^0,\varnothing^0)$\;\; $(\varnothing^0,	\begin{ytableau} *(white) \\ *(black)\end{ytableau}^0)$};
\draw (-0.2,-13) node[anchor=north]
	{$(\varnothing^0,	\begin{ytableau} *(white) & *(black)\end{ytableau}^0)$};
\draw (-10,-11.5) node[anchor=north]
	{ $(\varnothing^0,
	\begin{ytableau} *(white) & *(black) \\ *(black) \end{ytableau}^0)$
	$(\begin{ytableau} *(white) & *(black) \\ *(black) \end{ytableau}^0, \varnothing^0)$};
\draw (10,-14.8) node[anchor=north]
	{$(	\begin{ytableau} *(black) & *(white)\end{ytableau}^1,\varnothing^1)$};
\draw (10,-16.3) node[anchor=north]
	{$(	\begin{ytableau} *(black)\\ *(white)  \end{ytableau}^1,\varnothing^1)$\;\; $(\varnothing^1,	\begin{ytableau} *(black) \\ *(white) \end{ytableau}^1)$};
\draw (9.7,-17.8) node[anchor=north]
	{$(\varnothing^1,	\begin{ytableau} *(black) & *(white) \end{ytableau}^1)$};
\draw (-10,-14.8) node[anchor=north]
	{$(	\begin{ytableau} *(black) & *(white) & *(black) \end{ytableau}^1, \begin{ytableau} *(black)  \end{ytableau}^1)$};
\draw (-10,-16.3) node[anchor=north]
	{$(	\begin{ytableau} *(black)\\ *(white) \\ *(black) 		 \end{ytableau}^1,\begin{ytableau} *(black) 				\end{ytableau}^1)$\;\;
	$(\begin{ytableau} *(black) \end{ytableau}^1,				\begin{ytableau} *(black) \\ *(white) \\ *(black) \end{ytableau}^1)$};
\draw (-10.3,-18.3) node[anchor=north]
	{$(	\begin{ytableau} *(black) \end{ytableau}^1, \begin{ytableau} *(black) & *(white) & *(black) \end{ytableau}^1)$};
\draw (20,-22) node[anchor=north]
	{$(\begin{ytableau} *(white) \end{ytableau}^0,
	\begin{ytableau} *(white)\end{ytableau}^0)$};
\draw (-20,-21) node[anchor=north]
	{$\left(\begin{ytableau} *(white) & *(black) \\ *(black) \end{ytableau}^0,
	\begin{ytableau}  *(white) & *(black) \\ *(black) \end{ytableau}^0\right)$};
\draw (0,-16) node[anchor=north] {\dots\dots};
\draw (10,-22) node[anchor=north] {\dots\dots};
\draw (0,-22) node[anchor=north] {\dots\dots};
\draw (-10,-22) node[anchor=north] {\dots\dots};
%Vertical angles
\draw[very thin] (-18,-25) -- (-20,-20) -- (-22,-25);
\draw[very thin] (-3,-25) -- (-10,-5) -- (-17,-25);
\draw[very thin] (-8,-25) -- (0,1) -- (8,-25);
\draw[very thin] (3,-25) -- (10,-5) -- (17,-25);
\draw[very thin] (18,-25) -- (20,-20) -- (22,-25);
%Horizontal lines and values of L_0
\foreach \x/\xtext in {0/\Delta,1/\Delta+\frac12,2/\Delta+1, 3/\Delta+\frac32,4/\Delta+2}
	{\draw (24,-2-5*\x) node[anchor=west] {$L_0=\xtext$};
	\draw[very thin] (-24, -5-5*\x) -- (28, -5-5*\x);
	};
\foreach \x in {-4,-2,0,2,4}
	\draw (5*\x,-26) node[anchor=north] {$h_0=\x$};
\end{tikzpicture}
\caption{\small{The basis in $\calF^{(0,0)}(\kappa_1,\kappa_2)\oplus\calF^{(1,1)}(\kappa_1,\kappa_2)$. Pair of colored diagrams $(\lambda_1^\sigma,\lambda_2^\sigma)$ represents a vector  $J^{(2)}_{\vec{\lambda}^\sigma}$.
The interior of each angle corresponds to the representation $\calF \otimes \rmF_{2d} \otimes \sfV_\Delta \subset \calF^{(0,0)}(\kappa_1,\kappa_2)\oplus\calF^{(1,1)}(\kappa_1,\kappa_2)$.}} \label{fig_022}
\end{center}
\end{figure}

Taking the limit of formulas for the $P_{\vec{\lambda}}$ in terms of $a_n^{(1)}$ and $a_n^{(2)}$ (like \eqref{eq:r2:P}) and \eqref{eq:r2:wn}-\eqref{eq:r2:psir} one can find the first examples:

\begin{align}
J^{(2)}_{(\varnothing^0,\varnothing^0)}=|&P\rangle ,\quad \!J^{(2)}_{((1)^1,\varnothing^1)}=\left(G_{-1/2}{+}i\frac{Q{+}2P}{2}\chi_{-1/2}\right)|P\rangle, \quad \!J^{(2)}_{((1)^1,\varnothing^1)}=\left(G_{-1/2}{+}i\frac{Q{-}2P}{2}\chi_{-1/2}\right)|P\rangle, \notag \\
J^{(2)}_{((2)^0,\varnothing^0)}=&\left(-2L_{-1}-\frac{2i}{b} \chi_{-1/2}G_{-1/2}-\frac{i(Q+2P)}{2} w_{-1}-\frac{Q+2P}{2b}h_{-1}\right)|P\rangle,
 \notag \\ J^{(2)}_{((1,1)^0,\varnothing^0)}=&\left(-2L_{-1}-2bi  \chi_{-1/2}G_{-1/2}-\frac{i(Q+2P)}{2} w_{-1}-\frac{b(Q+2P)}{2}h_{-1}\right)|P\rangle,
\label{eq:r2:basis} \\ J^{(2)}_{(\varnothing^0,(2)^0)}=&\left(-2L_{-1}-\frac{2i}{b} \chi_{-1/2}G_{-1/2}-\frac{i(Q{-}2P)}{2} w_{-1}-\frac{Q{-}2P}{2b} h_{-1}\right)|P\rangle, \notag \\
J^{(2)}_{((2,2)^0,\varnothing^0)}=&\Bigl( 4L_{-1}^2-4G_{-3/2}G_{-1/2}-2Q(Q{+}2P)L_{-2}+ i4(Q+P)w_{-1}L_{-1}-\frac{(Q{+}P)(Q{+}2P)}2 w_{-1}^2 +  \notag \\ & +\frac{(Q{+}P)(Q{+}2P)}2 h_{-1}^2-2(Q{+}P)(Q{+}2P)\chi_{-3/2}\chi_{-1/2}-i Q(Q{+}P)(Q{+}2P)w_{-2}+ \notag \\ &  +i4(Q{+}P)h_{-1}\chi_{-1/2}G_{-1/2}\Bigr)|P\rangle; \notag
\end{align}

The basis $J^{(2)}_{\vec{\lambda}^\sigma}$ is orthogonal and normalization chosen such that:
\begin{align}\label{eq:r2:norm}
\langle J^{(2)}_{\vec{\lambda}^\sigma}|J^{(2)}_{\vec{\lambda}^{\sigma}}   \rangle=
\begin{cases}
\prod_{i,j=1}^2 N^{(2)}_{\lambda_i^\sigma,\lambda_j^{\tilde{\sigma}}}(P_i-P_j)  \quad &\text{if} \;\; |\vec{\lambda}| \equiv 0 \!\!\mod \,2 \\
\frac12 \prod_{i,j=1}^2 N^{(2)}_{\lambda_i^\sigma,\lambda_j^{\tilde{\sigma}}}(P_i-P_j)  \quad &\text{if} \;\; |\vec{\lambda}| \equiv 1 \!\!\mod \,2,
\end{cases}
\end{align}
where $(P_1,P_2)=(P,-P)$ and $N^{(2)}_{\lambda^\sigma, \mu^{\tilde{\sigma}}}(\alpha)$ defined in \eqref{eq:N(2)}.

In the remaining part of this section we will state some properties of this basis. For simplicity we will consider only the subspace $\calF \otimes \rmF_{0} \otimes \sfV_\Delta$ i.e subspace where $h_0=0$. In combinatorial terms it means that $$\Bigl(\sigma=0,\; \;  N_0(\vec{\lambda})=N_1(\vec{\lambda})\Bigr) \quad \text{or} \quad
\Bigl(\sigma=1,\;\; N_1(\vec{\lambda})=N_0(\vec{\lambda})+1\Bigr).$$
\secpart Elements of the form $J^{(2)}_{\lambda^\sigma,\varnothing^\sigma}$ can be given in explicit form. After bosonization of the basis elements (using sign $"+"$ in formula \eqref{eq:NSR:bos}) we have expressions like:
\begin{align}
&J^{(2)}_{((1)^0,\varnothing^0)} = \frac{i(Q+2P)}2\left( \chi_{-1/2}+\psi_{-1/2}\right)|P\rangle, \notag\\
&J^{(2)}_{((2)^0,\varnothing^0)} = -(Q+2P)\left( i(\frac12w_{-1}+c_{-1})+b^{-1}(\frac12h_{-1}+\chi_{-1/2}\psi_{-1/2})\right)|P\rangle, \notag\\
&J^{(2)}_{((1,1)^0,\varnothing^0)} = -(Q+2P)\left( i(\frac12w_{-1}+c_{-1})+b(\frac12h_{-1}+\chi_{-1/2}\psi_{-1/2})\right)|P\rangle. \notag
\end{align}
Recall that in deformed case basis vectors $P_{\lambda,\varnothing}=P_\lambda\otimes 1$ (see \ref{sub:Plambda})). So in the limit the basis vectors $J^{(2)}_{\lambda^\sigma,\varnothing^\sigma}$ are identified to the Uglov polynomials $J_\lambda$ of the first Heisenberg algebra $a_n^{(1)}$ up to some normalization factor which we demote by $\Omega_\lambda(P)$. This factor should be in agreement wtih the formula~\eqref{eq:r2:norm}. We conjecture $\Omega_\lambda(P)$ have the form
\begin{align}\label{eq:Omega}
\Omega_\lambda(P)=\prod_{s \in \lambda,\; i+j \equiv 0 \!\!\!\! \mod  2} (2P+ib^{-1}+jb).
\end{align}

\begin{Conj} \label{Conj:J:varnohing}
(1) Let $\sigma=0$, $N_0(\vec{\lambda})=N_1(\vec{\lambda})$. After bosonization we have:  \begin{align*}
J^{(2)}_{\lambda^\sigma,\varnothing^\sigma}=\Omega_\lambda(P) J^{(2)}_\lambda(a_k^{(1)}),
\end{align*}
where $J^{(2)}_\lambda$ defined in \eqref{eq:J(2):def},\eqref{eq:J(2):exmp}. The polynomial $J^{(2)}_\lambda$ can  also be rewritten in terms of two Heisenberg algebras $h_n^{(1)}$ and $w_n^{(1)}$ which can be found from $a_n^{(1)}$ as in \eqref{eq:h,w:a}:
\begin{align*}
w^{(1)}_n=a^{(1)}_{2n},\qquad \sum h^{(1)}_n z^{-2n}=\frac12 +\frac{(-1)^{\sigma+1}}4 \left(\exp(2\phi^{(1)})+ \exp(-2\phi^{(1)})\right).
\end{align*}
(2) Let $\sigma=1$, $N_1(\vec{\lambda})=N_0(\vec{\lambda)}+1$. After bosonization we have:  \begin{align*}
J^{(2)}_{\lambda^\sigma,\varnothing^\sigma}=\frac{-1}{\sqrt{2}}\Omega_\lambda(P) J^{(2)}_\lambda(a_k^{(1)}),
\end{align*}
where $J^{(2)}_\lambda$ defined in \eqref{eq:J(2):def},\eqref{eq:J(2):exmp}.
\end{Conj}
\noindent Operators $h_n^{(1)}$ and $w_n^{(1)}$ used in Conjecture \ref{Conj:J:varnohing} can be expressed by use of \eqref{eq:r2:wn}-\eqref{eq:r2:psir}:
\begin{align*}
w^{(1)}_n=\frac12 w_n+c_n,\qquad h_n^{(1)}=\frac12 h_n- \sum \chi_r\psi_{n-r}.
\end{align*}

One can use another free field representation which corresponds to the sign $"-"$ in formula \eqref{eq:NSR:bos}). Denote the corresponding generators by $\tilde{c}_n,\tilde{\psi}_r$. In this case the vectors $J^{(2)}_{\varnothing^\sigma, \lambda^\sigma}$ are identified with Uglov polynomials $J_{\lambda^\sigma}$ depending on $\tilde{a}_n^{(1)}$, with factor $\Omega_\lambda(-P)$.

In the case $Q=b+b^{-1}=0$ the situation simplifies more. It follows from \eqref{eq:NSR:bos} that in this case $\tilde{c}_k=-c_k$ and $\tilde{\psi}_r=-\psi_r$. Therefore $\tilde{a}_n^{(1)}=a_n^{(2)}$ and generic vectors $J_{\lambda_1^\sigma,\lambda_2^\sigma}$ become a product of two the Uglov polynomials $J_{\lambda_1^\sigma}(a_n^{(1)})\cdot J_{\lambda_2^\sigma}(a_n^{(2)})$. The analogous fact for the algebra $\calA(1,2)=\calH\oplus \Vir$ was given in  \cite{Belavin:2011js}.

The corresponding Heisenberg generators $h_n^{(2)}$ and $w_n^{(2)}$ read:
\begin{align*}
&w^{(2)}_n=a^{(2)}_{2n},\qquad \sum h^{(2)}_n z^{-2n}-\frac12 = \frac{(-1)^{\sigma+1}}4 \left(\exp(2\phi^{(2)})+ \exp(-2\phi^{(2)})\right);\\
&w^{(2)}_n=\frac12 w_n-c_n, \qquad h_n^{(2)}=\frac12 h_n+ \sum \chi_r\psi_{n-r}.
\end{align*}

\secpart It remains to take the limit of the vertex operator $\Phi(z)$ for the algebra $\calE_1(q,t)$. Since we consider the representation \eqref{eq:r2:rep} then we will take the limit of two vertex operators:
\begin{align*}
&\Phi_{0}(z) \colon \calF^{(0,0)}(\kappa_1,\kappa_2) \rightarrow \calF^{(0,0)}(\tilde{\kappa}_1,\tilde{\kappa}_2),\qquad \Phi_{0}(z) \colon \calF^{(1,1)}(\kappa_1,\kappa_2) \rightarrow \calF^{(1,1)}(\tilde{\kappa}_1,\tilde{\kappa}_2),
\\ &\Phi_{1}(z) \colon \calF^{(0,0)}(\kappa_1,\kappa_2) \rightarrow \calF^{(1,1)}(\tilde{\kappa}_1,\tilde{\kappa}_2),\qquad \Phi_{1}(z) \colon \calF^{(1,1)}(\kappa_1,\kappa_2) \rightarrow \calF^{(0,0)}(\tilde{\kappa}_1,\tilde{\kappa}_2).
\end{align*}
Similar to $r=1$ case norms \eqref{eq:conj:norm:level:m} and matrix elements \eqref{eq:DI:matrix} tend to 0 when $\tau \rightarrow 0$. Combinatorial considerations suggest that in the case $\sigma=\tilde{\sigma}$ we have:
$$\lim\limits_{\tau \rightarrow 0} \left(\frac{\langle  K_{\vec{\lambda}} |\Phi(z) |K_{\vec{\mu}}\rangle}{\sqrt{\langle  K_{\vec{\lambda}} K_{\vec{\lambda}}\rangle}\cdot \sqrt{\langle  K_{\vec{\mu}} K_{\vec{\mu}}}\rangle}\right)=
\begin{cases}
\neq 0 \quad & \text{if $h_0$ have the same eigenvalues on $J^{(2)}_{\vec{\lambda}^\sigma}$ and $J^{(2)}_{\vec{\mu}^{\tilde{\sigma}}}$}  \\
0, \quad &\text{otherwise} \;\;
\end{cases} $$
In the case $|\sigma-\tilde{\sigma}|=1$ we have:
$$\lim\limits_{\tau \rightarrow 0}\left(\frac{\tau\langle  K_{\vec{\lambda}} |\Phi(z) |K_{\vec{\mu}}\rangle}{\sqrt{\langle  K_{\vec{\lambda}} K_{\vec{\lambda}}\rangle}\cdot \sqrt{\langle  K_{\vec{\mu}} K_{\vec{\mu}}}\rangle}\right)=
\begin{cases}
\neq 0 \quad &\text{if $h_0$ have the same eigenvalues on $J^{(2)}_{\vec{\lambda}^\sigma}$ and $J^{(2)}_{\vec{\mu}^{\tilde{\sigma}}}$} \\
0, \quad &\text{otherwise}
\end{cases} $$
Therefore in the leading order of the limit of operators $\Phi_0(z)$ and $\Phi_1(z)$ we obtain the operator which commute with $h_0$. We suggest that this limit coincides with the vertex operator
\begin{align*}
 \mathcal{V}_{\alpha}(z) \cdot \Phi_{\alpha}^{\scriptscriptstyle{\textsf{NS}}}(z).
\end{align*}
Here $\mathcal{V}_{\alpha}$ is a standard rotated vertex operator on Heisenberg algebra $\mathcal{H}$:
\begin{align*}
\mathcal{V}_{\alpha}(z)=\exp\Big(i(\alpha-Q)\sum_{n=1}^{\infty}
\frac{w_{-n}z^{n}}{2n}\Big)\exp\Big(i\alpha\sum_{n=1}^{\infty}
\frac{w_{n}z^{-n}}{-2n}\Big)
\end{align*}
and $\Phi_{\alpha}^{\scriptscriptstyle{\textsf{NS}}}$ is the primary field of the \textsf{NSR} algebra with conformal
dimension $\Delta(\alpha)=\frac{1}{2}\alpha(Q-\alpha)$, $\Psi_{\alpha}^{\scriptscriptstyle{\textsf{NS}}}$ its super partner with the dimension $\Delta(\alpha)+1/2$. These operators can be defined by the commutation relations:
\begin{equation}
\begin{aligned}
&[L_{n},\Phi_{\alpha}^{\scriptscriptstyle{\textsf{NS}}}] =
(z^{n+1}\partial_{z}+(n+1)\Delta(\alpha)z^{n})\Phi_{\alpha}^{\scriptscriptstyle{\textsf{NS}}},\\
& [L_{n},\Psi_{\alpha}^{\scriptscriptstyle{\textsf{NS}}}] =
(z^{n+1}\partial_{z}+(n+1)(\Delta(\alpha)+1/2)z^{n})\Psi_{\alpha}^{\scriptscriptstyle{\textsf{NS}}},\\
&[G_{r},\Phi_{\alpha}^{\scriptscriptstyle{\textsf{NS}}}] =
z^{r+1/2}\Psi_{\alpha}^{\scriptscriptstyle{\textsf{NS}}},\\
&\{G_{r},\Psi_{\alpha}^{\scriptscriptstyle{\textsf{NS}}}\}=
(z^{r+1/2}\partial_{z}+(2r+1)\Delta(\alpha)z^{r-1/2})\Phi_{\alpha}^{\scriptscriptstyle{\textsf{NS}}},\end{aligned}
\end{equation}

\begin{Conj} \label{Conj:r2ME} Consider two colored partitions $\lambda^{\tilde{\sigma}}$, $\mu^{\sigma}$ such that $h_0$ gradings on $J^{(2)}_{\vec{\lambda}^{\tilde\sigma}}$ and  $J^{(2)}_{\vec{\mu}^{\sigma}}$ coincide.
Then matrix elements of the vertex operator $ \mathcal{V}_{\alpha}(z) \cdot \Phi_{\alpha}^{\scriptscriptstyle{\textsf{NS}}}(z)$ have completely factorized form:
\begin{align}\label{eq:r2ME}
\langle J^{(2)}_{\vec{\lambda}^{\tilde\sigma}} |\mathcal{V}_{\alpha}(z) \cdot \Phi_{\alpha}^{\scriptscriptstyle{\textsf{NS}}}(z)| J^{(2)}_{\vec{\mu}^{\sigma}}\rangle=
\begin{cases}
\prod_{i,j=1}^2 N^{(2)}_{\lambda_i^{\tilde{\sigma}},\mu_j^{\sigma}}(\alpha+P_i-P_j')  \quad &\text{if} \;\; \tilde{\sigma}=0
\\ -\prod_{i,j=1}^2 N^{(2)}_{\lambda_i^{\tilde{\sigma}},\mu_j^{\sigma}}(\alpha+P_i-P_j') \quad &\text{if} \;\; \tilde{\sigma}=1, \sigma=0
\\ \frac12 \prod_{i,j=1}^2 N^{(2)}_{\lambda_i^{\tilde{\sigma}},\mu_j^{\sigma}}(\alpha+P_i-P_j') \quad &\text{if} \;\; \tilde{\sigma}=1, \sigma=1,
\end{cases}
\end{align}
where $N^{(2)}_{\lambda^{\tilde{\sigma}}, \mu^{\sigma}}(\alpha)$ defined in \eqref{eq:N(2)}.
\end{Conj}
This Conjecture was checked up to the level 2. These checks provides very nontrivial evidence that at the limit $q,t \rightarrow -1$ of $\calE_1(q,t)$ we will have primary fields of the Neveu--Schwarz--Ramond algebra.

\section{Singular vectors}

The Uglov symmetric polynomials are natural generalization of the Jack symmetric polynomials. The singular vectors of the Virasoro algebra have the remarkable description in terms of Jack polynomials \cite{MimachiYamada1995,AMOS:1995}. The Uglov polynomials have the similar application to the representation theory of the Neveu--Schwarz--Ramond algebra.

We assume that parameter $b$ is generic. Let $m,n \in \Z$ such that $m-n \equiv 0 \!\! \mod \, 2$, and
\begin{align}
P_{m,n}=- (mb^{-1}+nb)/2.
\end{align}
Then for $P=P_{m,n}$ there exists operator $D_{m,n}$ such that vector $D_{m,n}|P_{m,n}\rangle$ is singular i.e.
$$L_k D_{m,n}|P_{m,n}\rangle=G_r D_{m,n}|P_{m,n}\rangle=0, \; k,r>0 \qquad L_0D_{m,n}|P_{m,n}\rangle=(\Delta_{m,n}{+}\frac{mn}2)D_{m,n}|P_{m,n}\rangle,$$
where $\Delta(P)=\frac{1}{2}(Q^2/4-P^2)$. The vector $D_{m,n}|P_{m,n}\rangle$ generates a submodule in $\sfV_{\Delta_{m,n}}$. For $P=P_{m,n}$ the operator $D_{m,n}$ is unique up to normalization, for other values of $P$ the Verma module $\sfV_\Delta$ is irreducible \cite{Kac:1979,Astashkevich}.

The first examples of the singular vectors $D_{m,n}|P_{m,n}\rangle$ have the form:
\begin{align*}
&D_{1,1}|P_{1,1} \rangle = G_{-1/2}|P_{1,1}\rangle,  \\
&D_{3,1}|P_{1,3}\rangle  = (L_{-1}G_{-1/2}+b^{2}G_{-3/2})|P_{1,3}\rangle,  \\
&D_{2,2}|P_{2,2} \rangle = (L_{-1}^{2}+\frac{1}{2}Q^{2}L_{-2}-G_{-3/2}G_{-1/2})|P_{2,2}\rangle.
\end{align*}
Expressing these $D_{m,n}$ through the free field generators $\tilde{c}_k, \tilde{\psi}_k$ (using sign $"-"$ in formula \eqref{eq:NSR:bos}) we get:
\begin{align*}
&D_{1,1}|P_{1,1} = -iQ\tilde{\psi}_{-1/2}|P_{1,1}\rangle, \\
&D_{3,1}|P_{1,3}\rangle = -Q(Q+2b)(\tilde{c}_{-1}\tilde{\psi}_{-1/2}+ib\tilde{\psi}_{-3/2})|P_{1,3}\rangle, \\
&D_{2,2}|P_{2,2} \rangle =  -2Q^{2}(\tilde{c}_{-1}^{2}+\frac{i}{2}Q \tilde{c}_{-2}-2\tilde{\psi}_{-3/2}\tilde{\psi}_{-1/2})|P_{2,2}\rangle.
\end{align*}
\begin{Conj}\label{Conj:Sing}
Denote by $(m^n)$ the rectangular Young diagram $(\underbrace{m,m,\dots,m}_n)$. The operators $D_{m,n}$ coincide (up to normalization) with the Uglov polynomials $J^{(2)}_{(m^n)}$ defined in \eqref{eq:J(2):def},\eqref{eq:J(2):exmp} with identification:
\begin{align}\label{rq:ferm}
\tilde{c_n}=\frac12a_{2n},\qquad \sum_r \tilde{\psi}_rz^{-2r}=\frac{i}{2\sqrt{2}}\left(\exp(\phi^-+2\phi^+)- \exp(-\phi^--2\phi^+) \right),
\end{align}
where $\phi^{+}(z)=\sum_{n \in \Z_>0} \dfrac{a_{2n+1}}{-2n-1} z^{-2n-1}$, $\phi^{-}(z)=\sum_{n \in \Z_<0} \dfrac{a_{2n+1}}{-2n-1} z^{-2n-1}.$
\end{Conj}
We have checked this Conjecture up to the level $9/2$. This Conjecture  is one of the main results of the paper.

The singular vectors related to the basis $J^{(2)}_{\lambda_1^\sigma,\lambda_2^\sigma}$ as follows. If we put $P=P_{m,n}$, then the factor $\Omega_{(m^n)}(P_{m,n})$ \eqref{eq:Omega} equals to 0. Therefore, the vector  $J^{(2)}_{(m^n)^\sigma,\varnothing^\sigma}$, (where $\sigma \equiv n \pmod 2$) vanishes after $c_n, \psi_r$ free field realization due to Conjecture \ref{Conj:J:varnohing}. Therefore, the vector $J^{(2)}_{(m^n)^\sigma,\varnothing^\sigma}$ belongs to the kernel of the natural map from the Verma module to the Fock module. Then the vector $J^{(2)}_{(m^n)^\sigma,\varnothing^\sigma}$ is singular for the algebra $\calA(2,2)=\calH\oplus\Asl(2)_2\oplus\NSR$.

The algebra $\calH$ has no singular vectors. For the algebra  $\Asl(2)_2$ we consider only integrable representations which also don't have singular vectors. Therefore $J^{(2)}_{(m^n),\varnothing}$ is singular vector for the $\NSR$ algebra. Hence if we put $P=P_{m,n}$ then all terms depending on $w_n,h_n,\chi_r$ vanish and the vector will depend only on $L_n, \psi_r$. One can easily see this on the examples \eqref{eq:r2:basis}. If we use the second bosonization we will get the vector depending only on $\tilde{c}_n,\tilde{\psi}_r.$  The Conjecture \ref{Conj:Sing} means that this vector can be described in terms of the Uglov polynomials.

It is worth to note that in the recent paper \cite{Desrosies:2012} Desrosiers, Lapointe and Mathieu proposed an another expression for the singular vectors $D_{m,n}$. In the formula in loc.~cit. $D_{m,n}$ become a linear combination of the \emph{superJack polynomials} contrary to our result where the only one \emph{Uglov polynomial} $J^{(2)}_{(m^n)}$ is used. From the other hand expressions from loc.~cit. do not involve nonpolynomial change of variables as in Conjecture \ref{Conj:Sing}.

\section{Concluding remarks}
\sectpart  In the paper we dealt only with the $p=2$ case but the methods are quite general. It is natural to expect that in the limit $q,t \rightarrow  \sqrt[p]{1}$ of the representations of the quantum toroidal $\mathfrak{gl}(1)$ algebra $\calE_1(q,t)$ on get the representation of the algebra $\calA(r,p)$. For example limit of the matrix elements \eqref{eq:conj:norm:level:m} for $\calE_1(q,t)$ coincides with the matrix elements \eqref{eq:matrix:color} for the $\calA(r,p)$. The rank $p$ Uglov polynomials form a natural basis in the representation of the algebra $\calA(1,p)=\calH\oplus \Asl(p)_1$ \cite{Uglov 2}.

It would be interesting to give a mathematically rigorous proof of our construction of the algebra $\calA(r,p)$. Note that this limit construction should give principal (Lepowsky-Wilson) construction of this algebra. It is expected that this algebra is a subalgebra of more general Yangian algebra.

It worth to note some relevant works. Ten years ago Hara et al \cite{Hara:2000} considered the limit of the deformed Virasoro algebra where $q \rightarrow 1$, $t \rightarrow -1$ and observed in the limit the Lepowsky--Wilson realization of Verma module of the algebra $\Asl(2)$. This limit can be considered as the limit of the $r=2$ representations of the $\calE_1(q,t)$. From the geometric side of the AGT relation the limit algebra corresponds to the equivariant integration on the affine Laumon spaces. In terms of the gauge theory this corresponds to the presence of the surface operators~\cite{Alday:2010vg}.

%Michael Lashkevich and Yaroslav Pugai used more general root of unity limits of $\calE_1(q,t)$ in the recent study of formfactors \cite{Lashkevich}

Note also that Taro Kimura used the limit when $q,t$ go to the root of 1 in the matrix model which corresponds to the instanton counting on the ALE space \cite{Kimura} and more general toric singularities \cite{Kimura2}.

\sectpart In the $r=2$ case we have constructed only the vertex which doesn't depend on the $\Asl(2)_2$ part. In particular the equations \eqref{eq:r2ME} do not determine the basis $J^{(2)}_{\vec{\lambda}^{\tilde\sigma}}$, therefore we used the deformed algebra. It would be interesting to find the formulas similar \eqref{eq:r2ME} for more general vertex operators.

One of the natural approaches is to take the Ramond sector into consideration. It is expected that our construction works in the Ramond representation of the algebra $\calA(2,2)=\calH\oplus\Asl(2)_2\oplus\NSR$ and the resulting basis factorizes the vertex operator which maps Neveu--Schwarz representation to the Ramond one and back. This vertex should have the form like $\calV_{\alpha}\cdot \Phi_{1/2} \cdot R$. Here $R$ is the Ramond primary field and $\Phi_{1/2}$ corresponds to the highest weight vector in $\calL_{0,2}$ by the operator-state correspondence.
This operator doesn't commute with the $h_0$ (on the instanton side changes the first Chern class $c_1$). Some checks for this vertex were made in \cite{Baur}.

\sectpart Recall that the AGT relation suggests that the Nekrasov partition functions i.e. the equivariant integrals on the moduli space of instantons coincide with the conformal blocks. These conformal blocks can be computed by use of matrix elements of the vertex operators. The formulae for the matrix elements given in Propositions \ref{Prop_ME_0}, \ref{Prop_ME_1} and Conjecture \ref{Conj:r2ME} coincides with the geometrically calculated values. 

For example for the algebra $\calA(1,2)=\calH\oplus\Asl(2)_1$ we have natural vertex operators $\calV_\alpha(z)$, $\calV_\alpha(z) \mathcal{W}_+(z^2)$, $\calV_\alpha(z) \mathcal{W}_-(z^2)$ and from the results of section 2 follows that  
\begin{multline*}
^0 \langle \tilde{\kappa}| \calV_{\beta}(w) \calV_{\alpha}(z) |\kappa \rangle^0  = \left(1{-}z^2/w^2\right)^{-\frac{\beta(Q-\alpha)}2}= \!\!\!\!\! \sum_{\lambda^{0}\!,\;\; d(\lambda^{0})=0} \!\!\!\!\!\!\! \frac{ N^{(2)}_{\varnothing^{0}, \lambda^{0}}(\beta) \cdot N^{(2)}_{\lambda^{0}, \varnothing^{0}}(\alpha)}{N^{(2)}_{\lambda^{0}, \lambda^{0}}(0)} (z/w)^{2|\lambda|}, \\
^0 \langle \tilde{\kappa}| \calV_{\beta}(w)\mathcal{W}_-(w^2) \calV_{\alpha}(z)\mathcal{W}_+(z^2)  |\kappa\rangle^0 = \left(1{-}z^2/w^2\right)^{-\frac{\beta(Q-\alpha)}2-\frac12} =\!\!\!\!\!\sum_{\lambda^{1}\!,\;\; d(\lambda^{1})=-1} \!\!\!\!\!\!\! \frac{ N^{(2)}_{\varnothing^{0}, \lambda^{1}}(\beta) \cdot N^{(2)}_{\lambda^{1}, \varnothing^{0}}(\alpha)}{N^{(2)}_{\lambda^{1}, \lambda^{1}}(0)} (z/w)^{2|\lambda|-1}, 
\end{multline*}
where $N^{(2)}_{\lambda^{\tilde{\sigma}}, \mu^{\sigma}}(\alpha)$ defined in \eqref{eq:N(2)}. The additional factor $\left(1{-}z^2/w^2\right)^{-\frac12}$ equals to the nontrivial conformal block for $\Asl(2)_1$.

Using the basis and the matrix elements given in section 3 one can write formulae for conformal blocks for the algebra $\calA(2,2)=\calH\oplus\Asl(2)_2\oplus\NSR$. The contribution of the $\calH\oplus\Asl(2)_2$ can be factored out and we get formula for the $\NSR$ conformal block as in \cite{BBB:2011tb}.

Note that there exists the other formula for $\NSR$ (better to say $\calA(2,2)$) confromal block \cite{Bonelli:2011}. This formula follows
from the properties of the other basis (constructed in \cite{BBFLT:2011})

\section{Acknowledgments}

We thank  M. Lashkeich, A. Litvinov, N. Nekrasov, V. Pasquier, Y. Pugai, D. Serban, J. Shiraishi, L. Spodyneiko  and especially B. Feigin for interest to our work and discussions.  We are grateful for organizers of the workshop on the AGT Conjecture, held in Bethe Forum, Bonn, March 2012 and especially R. Flume for hospitality and stimulating scientific atmosphere. M.B. thanks Simons Center for Geometry and Physics where this work was finished for their hospitality during his visit.

This work was supported by RFBR grants No.12-01-00836-a, 12-02-01092-a, 12-01-31236-mol\_a, 12-02-33011-mol\_a\_ved and by the Russian Ministry of Education and Science under the grants 2012-1.5-12-000-1011-012, contract No.8528, 2012-1.1-12-000-1011-016, contract No.8410 and agrremennt No.14.A18.21.2027.

\Appendix

\section{Partitions} \label{AppendixYoung}

We basically follow \cite{Macdonald Book} for the notations. A partition $\lambda$ is a series of nonnegative integers $\lambda=(\lambda_1,\lambda_2,\ldots)$ such that $\lambda_1\ge\lambda_2\ge\cdots$ with finitely many nonzero entries, $|\lambda|  \seteq \sum_{i\geq 1} \lambda_i$. If $\lambda_l>0$ and $\lambda_{l+1}=0$,
we write $\ell(\lambda) \seteq l$ and call it the length of $\lambda$. We identify the partition and the corresponding Young diagram.

The conjugate partition of $\lambda$ is denoted by $\lambda'$ which corresponds to
the transpose of the diagram $\lambda$. The empty partition is denoted by $\varnothing$.
The dominance ordering is defined by $\lambda\ge\mu$ $\Leftrightarrow$
$|\lambda|=|\mu|$ and $\sum_{k=1}^i \lambda_k \ge \sum_{k=1}^i \mu_k$ for all $i=1,2,\ldots$.

In Section \ref{seqp=2} and later we use the colouring of Young diagrams in two colors, labeled by residues modulo $2$: the box $s\in \lambda$ with coordinates $(i,j)$ has color $i-j+\sigma$. Here $\sigma=0,1$ is the color of the angle. We will denote such colored Young diagram by $\lambda^{\sigma}$ in order to stress the coloring. By $N_i(\lambda^\sigma)$ we denote the number of boxes of color $i$, $i=0, 1$. Let $d(\lambda^\sigma)=N_0(\lambda^\sigma)-N_1(\lambda^\sigma)$. The following lemmas are standard.

\begin{lemma}
Let $\sigma=0$. There is a smallest partition $\mu$ with $d(\mu)=d$. This partition consists of $2d^2-d$ boxes and has a ``triangular'' form with edge length  $2|d|$ for $d \leq 0$ and $2d-1$ for $d>0$.

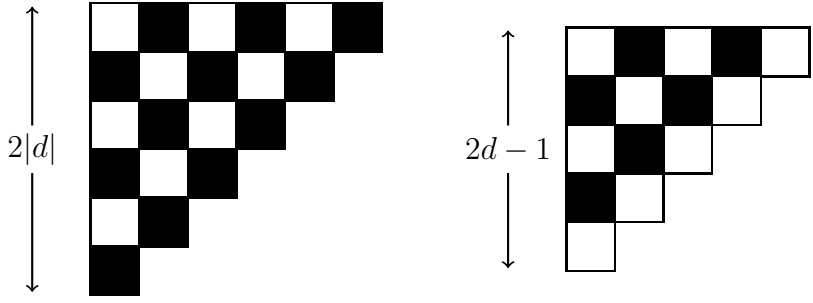
\ytableausetup{mathmode, boxsize=1.5em}
\begin{center}
\begin{tikzpicture}[x=1.5em, y=1.5em]
\draw[thick,->] (0,-0.5) -- (0,-3) ;
\draw (0,0) node {$2|d|$};
\draw[thick,->] (0,0.5) -- (0,3);
\draw (1,0) node[anchor=west]
	{\begin{ytableau}
	*(white) & *(black) & *(white) & *(black) & *(white) & *(black)\\
	*(black) & *(white) & *(black) & *(white) & *(black) \\
	*(white) & *(black) & *(white) & *(black)\\
	*(black) & *(white) & *(black) \\
	*(white) & *(black)\\
	*(black)
	\end{ytableau}};
\draw[thick,->] (10,-0.5) -- (10,-2.5) ;
\draw (10,0) node {$2d-1$};
\draw[thick,->] (10,0.5) -- (10,2.5);
\draw (11,0) node[anchor=west]
	{\begin{ytableau}
	*(white) & *(black) & *(white) & *(black) & *(white) \\
	*(black) & *(white) & *(black) & *(white) \\
	*(white) & *(black) & *(white) \\
	*(black) & *(white) \\
	*(white) \\
	\end{ytableau}};
\end{tikzpicture}
\end{center}
If $\sigma=1$ then the smallest partition $\mu$ with $d(\mu)=d$ has $2d^2+d$ boxes.
\end{lemma}

Such minimal diagram is called 2-core (\cite[Sec 1.1 Ex. 8]{Macdonald Book}). For any partition $\lambda$ its 2-core is denoted by $\widetilde{\lambda}$ and can be obtained as follows. Remove the rectangle $1 \times 2$ (or $2 \times 1 $) from the diagram of $\lambda$ in such a way that what remains is the diagram of a partition, and continue removing such rectangles in this way as long as possible. The result of this process is the 2-core $\widetilde{\lambda}$ of $\lambda$ and the result is independent of the sequence of removals.

\begin{lemma}
The number of partitions $\lambda$ such that $d(\lambda)=d$ and $|\lambda|-|\widetilde{\lambda}|=2n$ equals to the number of pairs of partitions $(\mu_1,\mu_2)$ such that $|\mu_1|+|\mu_2|=n$.
\end{lemma}
This Lemma has a bijective proof (see e.g. \cite[Sec 1.1 Ex. 8]{Macdonald Book}).  In terms of generating functions we have that
\begin{align}\label{eqchi^(1)}
\chi^{(1)}_{d,\sigma}(q):=\sum_{\lambda^\sigma,d(\lambda^\sigma)=d} x^{|\lambda|/2}=
x^{\frac{2d^2-(-1)^\sigma d}{2}}\prod_{k \in \Z_{>0}}\frac1{(1-x^k)^2}
=x^{\frac{2d^2-(-1)^\sigma d}{2}} \chi_B^2.
\end{align}
Note that this combinatorial fact has algebraic explanation:~\eqref{eqJ_basis_F0}, \eqref{eqJ_basis_F1}.

By $a_{\lambda}(s)$ and $l_{\lambda}(s)$ we denote the arm and the leg lengths of any box $s=(i,j)$: 
\begin{align} \label{eqarm_leg}
a_\lambda(s)=\lambda_j-i,\qquad \qquad l_{\lambda}(s)=\lambda'_i-j.
\end{align}
Note that box $s$ can be outside Young diagram of $\lambda$.
\begin{lemma}
Let $\lambda^{\lozenge}=\{s \in \lambda \mid a_\lambda(s)+l_\lambda(s)+1\equiv 0 \!\! \mod \, 2\}$. Then $|\lambda^\lozenge|=\dfrac{|\lambda|-|\widetilde{\lambda}|}{2}$.
\end{lemma}
This Lemma can be proved by induction. In the basis one observes that if $\lambda$ has triangular form ($\lambda=\widetilde{\lambda}$) then $\lambda^\lozenge=\varnothing$. In the inductive step one observes that after removing rectangle $1\times 2$ the $|\lambda^\lozenge|$ decreases by 1.
\section{Symmetric polynomials}
\subsection{Macdonald polynomials} \label{Appendix_Macdonald_polyn}
\secpart
We basically follow \cite{Macdonald Book} for the notations for the symmetric polynomials: $m_\lambda$~--- monomial symmetric functions, $p_\lambda=p_{\lambda_1}\dots p_{\lambda_n}$ and $p_k$ power-sum symmetric functions. If $\lambda$ has $m_i=m_i(\lambda)$ parts equal to $i$, then write
\begin{align*}
z_\lambda=(1^{m_1}2^{m_2}\cdots)m_1!m_2!\cdots.
\end{align*}
We use notation
\begin{align}
n(\lambda) \seteq \sum_{i\geq 1}(i-1)\lambda_i=\sum_{j=1}^{l(\lambda')}\binom{\lambda_j'}{2}. \label{eqn_lambda}
\end{align}
\secpart The Macdonald symmetric functions are denoted by $P_\lambda(q,t)$. These polynomials are defined by two properties:
\begin{itemize}
\item  The transition matrix between the basis $P_\lambda(q,t)$ and the basis $m_\lambda$ is upper unitriangular
\begin{align} P_\lambda(q,t)=m_\lambda+\sum_{\mu <\lambda} u_{\lambda,\mu}(q,t)m_\mu, \label{eqMacdonald_in_m}
\end{align}
where the summation is over partitions $\mu< \lambda$ in the dominance order.
\item The polynomials $P_{\lambda}(q,t)$ are orthogonal under the scalar product:
\begin{align} \langle p_\lambda,p_\mu\rangle_{q,t}=\delta_{\lambda,\mu}z_{\lambda} \left(\prod_{i=1}^{l(\lambda)} \frac{1-q^{\lambda_i}}{1-t^{\lambda_i}}\right). \label{eqMacdonald_produc}
\end{align}
\end{itemize}
Note that the basis $P_\lambda(q,t)$ is not orthonormal and the norms equal:
\begin{align*} \langle P_\lambda(q,t),P_\lambda(q,t)\rangle_{q,t}= \prod_{s \in \lambda}b_\lambda(s;q,t)^{-1}=\prod_{s \in \lambda}\frac{1-q^{a(s)+1}t^{l(s)}}{1-q^{a(s)}t^{l(s)+1}},
\end{align*}
where $a_\lambda(s)$, $l_\lambda(s)$ are the arm and the leg length \eqref{eqarm_leg}.

\secpart For $\mu \subset \lambda$ the skew partition $\lambda/\mu$ is called the horizontal strip if the sequences $\lambda$ and $\mu$ are interlaced, in the sense that $\lambda_l\geq \mu_l \geq \lambda_2\geq \mu_2 \geq \dots$ A tableau $T$ of shape $\lambda$ is a sequence of partition
\begin{equation*}
\varnothing=\lambda^{(0)}\subset \lambda^{(1)} \subset \ldots \subset  \lambda^{(r)}=\lambda
\end{equation*}
such that for each $i$ the skew diagram $\lambda^{(i)}/\lambda^{(i-1)}$ is a horizontal strip. The sequence $\mu=|\lambda^{(1)}/\lambda^{(0)}|,|\lambda^{(2)}/\lambda^{(1)}|,\dots,|\lambda^{(r)}/\lambda^{(r-1)}|$ is called the weight of $T$.

The coefficients $u_{\lambda,\mu}(q,t)$ have an explicit combinatorial formula:
$$u_{\lambda,\mu}(q,t)=\sum_{T}\psi_T(q,t),$$
where the summation goes through tableaux of shape $\lambda$ and weight $\mu$. For partitions $\lambda$ and $\mu$ such that $\mu \subset \lambda$ and $\lambda/\mu$ is a horizontal strip, let $R_{\lambda/\mu}$ denotes the union of the rows of $\mu$ that intersect which intersects $\lambda/\mu.$ Then
\begin{align}
\psi_{T}(q,t) = \prod_{i=1}^r \psi_{\lambda^{(i)}/\lambda^{(i-1)}}(q,t),\qquad \text{where}\qquad \psi_{\lambda/\mu}(q,t) = \prod_{s \in R_{\lambda/\mu}} \frac{b_{\mu}(s;q,t)}{b_{\lambda}(s;q,t)}\label{eqpsi_T}.
\end{align}
Note that in notation $R_{\lambda/\mu}$ we follow e.g. \cite{Uglov 2} rather than \cite{Macdonald Book}, where the index set denoted by $R_{\lambda/\mu}-C_{\lambda/\mu}$. This is just notation difference.

\secpart For each partition $\lambda$ define
\begin{align*}
c_\lambda(q,t)=\prod_{s\in\lambda}(1-q^{a(s)}t^{l(s)+1}),\\
J_\lambda(q,t)=c_{\lambda}(q,t)P_\lambda(q,t).
\end{align*}
From the Macdonald conjectures (proved by Haiman) follows that coefficients $v_{\lambda \mu}$ in the expansion
\begin{align*}
J_\lambda(q,t)=\sum_{\mu \leq \lambda} v_{\lambda,\mu}(q,t)m_\mu,
\end{align*}
are polynomials on $q,t$ with the integer coefficients. The basis $J_\lambda(q,t)$ is orthogonal and norms are equal:
\begin{align*}
\langle J_\lambda(q,t),J_\lambda(q,t)\rangle_{q,t}=\prod_{s \in \lambda}(1-q^{a(s)}t^{l(s)+1})(1-q^{a(s)+1}t^{l(s)}).
\end{align*}

\subsection{Uglov polynomials} \label{Appendix_Uglov_polyn}

\secpart Let
\begin{align*}
q= \omega_{p} e^{-\tau \varepsilon_{2}}, \quad  t=
\omega_{p}&e^{\tau
\varepsilon_{1}}, \qquad \omega_{p} = e^{i\frac{2\pi}{p} },\qquad \tau\rightarrow\infty.
\end{align*}
The limit of the Macdonald polynomials $J_\lambda(q,t)$ and $P_\lambda(q,t)$ is called \emph{Uglov polynomials of rank $p$} $J_{\lambda}^{(\alpha, p)}$ and $P_{\lambda}^{(\alpha, p)}$, where $\alpha=-\eps_2/\eps_1$. In the case $p=1$ these polynomials are standard Jack polynomials. The notation $\alpha$ is a standard for the Jack polynomials and $\eps_1, \eps_2$ are standard for instanton counting. Denis Uglov in \cite{Uglov 1} used the term \emph{Jack($\mathfrak{gl}_p$) polynomials}. In this appendix we list some properties of Uglov polynomials.

\secpart The first question is the existence of the limit and linear independence of the Uglov polynomials. For $P^{(\alpha, p)}_\lambda$ we can take the limit in formula \eqref{eqMacdonald_in_m}
\begin{equation*}
P^{(\alpha, p)}_\lambda=m_\lambda+\sum_{\mu<\lambda}u^{(\alpha,p)}_{\lambda\mu}m_\mu,\quad \text{where}\quad u^{(\alpha)}_{\lambda\mu}=\sum_{T}\psi_T^{(\alpha, p)}
\end{equation*}
 sum as above goes over tableaux $T$ of shape $\lambda$ and weight $\mu$. Functions $\psi_T$, $\psi_{\lambda/\mu}$ defined as in \eqref{eqpsi_T} and
\begin{equation*}
b_{\lambda}^{(\alpha, p)}(s)=\left\{ \begin{aligned}
& \frac{(l_\lambda(s)+1)+\alpha a_\lambda(s)}{l_\lambda(s)+\alpha(a_\lambda(s)+1)} \quad \text{if}\quad  a_\lambda(s)+l_\lambda(s)+1\equiv 0 \bmod p, \\
& 1, \quad \quad \quad \text{otherwise}.\\
\end{aligned}\right.
\end{equation*}
The last fraction can also be rewritten as $\dfrac{\eps_1(l_\lambda(s)+1)-\eps_2a_\lambda(s)}{\eps_1 l_\lambda(s)-\eps_2(a_\lambda(s)+1)}$.

Therefore for generic $\alpha$ the limit polynomials $P^{(\alpha, p)}_\lambda$ exists and form a basis in the space of symmetric polynomials. The basis $P^{(\alpha, p)}_\lambda$ is orthogonal under the limit scalar product
\begin{align*}
\langle p_\lambda,p_\mu\rangle_{\alpha,p}=\delta_{\lambda,\mu}z_{\lambda}  \alpha^{|\{i \mid \lambda_i \equiv 0 \!\!\!\! \mod p \}|}.
\end{align*}
The norms of $P^{(\alpha, p)}_\lambda$ equal
\begin{align*} \langle P_\lambda^{\alpha,p},P_\lambda^{\alpha,p}\rangle_{\alpha,p}=\prod_{s \in \lambda}b_\lambda^{\alpha,p}(s)^{-1}.
\end{align*}

\secpart We will need integer normalization of the Uglov polynomials. It easy to see that for $\tau \ll 1$
\begin{align*}
c_\lambda(q,t) \sim \tau^{|\lambda^{\lozenge}|}\prod_{s\in\lambda-\lambda^{\lozenge}}\left(1-\omega_p^{a_\lambda(s)+l_\lambda(s)+1}\right)
\prod_{s\in \lambda^{\lozenge}}\Bigl(\eps_1l_\lambda(s)+\eps_1-\eps_2 a_\lambda(s)\Bigr),
\end{align*}
where by $\lambda^{\lozenge}=\{s \in \lambda \mid a_\lambda(s)+l_\lambda(s)+1\equiv 0 \!\! \mod \, p\}$. So the polynomials $J^{(\alpha, p)}_\lambda$ can be defined by
\begin{align}
\!\!\!\!J^{(\alpha, p)}_\lambda\!=\!\lim_{\tau \rightarrow \infty}\left(\frac{J_\lambda(q,t)} {\tau^{|\lambda^{\lozenge}|}\eps_1^{\lambda^{\lozenge}} \prod\limits_{s\in\lambda-\lambda^{\lozenge}} \!\!\left(1{-}\omega_p^{a_\lambda(s)+l_\lambda(s)+1}\right)} \!\right)= P^{(\alpha, p)}_\lambda\prod_{s\in \lambda^{\lozenge}}(l_\lambda(s)+1-\alpha a_\lambda(s)).
\end{align}
Then polynomials $J^{(\alpha, p)}_\lambda$ don't vanish. They form an orthogonal basis with norms:
\begin{align} \label{eqNorm_Uglov}
\langle J^{(\alpha, p)}_\lambda,J^{(\alpha, p)}_\lambda\rangle_{\alpha,p}=\prod_{s \in \lambda^{\lozenge}}(\eps_1l_\lambda(s)+\eps_1-\eps_2 a_\lambda(s))(\eps_1l_\lambda(s)-\eps_2 a_\lambda(s)-\eps_2).
\end{align}
The coefficients of $J^{(\alpha, p)}_\lambda$ on the decomposition on monomial basis are polynomials in $\alpha$.

\secpart Let us give some examples of $J_{\lambda}^{(\alpha,p)}$. Here $p=2$ (case considered in the main part of the paper). And polynomials are given in the decomposition on $p_\lambda$ basis.
\begin{align}
&J^{(\alpha, 2)}_{(1)}=p_1, \quad J^{(\alpha, 2)}_{(2)}=- p_2-\alpha p_1^2, \quad  J^{(\alpha, 2)}_{(1,1)}= p_2 - p_1^2, \quad J^{(\alpha, 2)}_{(2,1)}=-\frac{1}{3}p_3+\frac13 p_{1,1,1},  \notag \\
&J^{(\alpha, 2)}_{(3)}=-\frac{2}{3}\alpha p_3- p_2 p_1  -\frac{1}{3} \alpha p_1^3, \quad
J^{(\alpha, 2)}_{(1,1,1)}=-\frac{2}{3}  p_3+ p_2 p_1- \frac{1}{3} p_{1}^3,\notag \\
&J^{(\alpha, 2)}_{(4)}=2 \alpha p_4 +  p_{2}^2 +2\alpha  p_{2}p_1^2 +\frac{8}{3} \alpha^2 p_3p_1 +\frac{1}{3} \alpha^2 p_{1}^4  \notag \\
&J^{(\alpha, 2)}_{(3,1)}=-2\alpha p_4+\frac{2(1-\alpha)\alpha}3 p_3p_1- p_2^2 +(1+\alpha) p_2 p_1^2 + \frac{(1+\alpha)\alpha}3 p_1^4 \label{eq:App:Ugl} \\
&J^{(\alpha, 2)}_{(2,2)}=(\alpha-1)p_4-\frac{4}{3}  \alpha p_3p_1+p_2^2-\frac{1}{3}\alpha p_1^4 \notag
\\&J^{(\alpha, 2)}_{(2,1,1)}=2 p_4 -\frac{2(1-\alpha)}{3} p_3p_1-p_2^2 -(1+\alpha)p_2p_1^2+\frac{(2+\alpha)}{3} p_{1}^4 \notag
\\&J^{(\alpha, 2)}_{1,1,1,1}=-2 p_4+p_2^2 +\frac{8}{3}  p_3p_1 -2 p_2p_1^2 +\frac{1}{3}p_1^4. \notag
\end{align}

\section{Fermion currents} \label{AppOpeSta}
The representations considered in the main text should have a structure of the Vertex-operator algebra. It means that for any vector in the representation corresponds a local field i.e. the power series of operators.

We mainly follow \cite{Fateev:1985mm} in the notations. The highest weight vector of the representation $\calL_{0,1}$ is denoted as $\Phi_0^0$ and corresponds to the identity operator $I$. On the top of the representation $\calL_{1,1}$ we have two vectors (see figure \ref{fig_11}) which are denoted by $\Phi_{1/2}^{1/2}$ and $\Phi_{-1/2}^{1/2}$ and correspond to the local fields:
\begin{align*}
\Phi_{1/2}^{1/2}(z)=\exp(\varphi(z)), \quad \Phi_{-1/2}^{1/2}(z)=\exp(-\varphi(z)),
\end{align*}
where $\varphi(z)$ is defined in \eqref{eq:varphi}. Due to Proposition \ref{Prop:W} this formula can be rewritten in terms of the field $\phi(z)$ defined in \eqref{eq:phi}. Namely:
\begin{align*}
&\Phi_{1/2}^{1/2}(z^2)=\exp(\varphi(z^2)) = \frac{z^{1/2}}2\Bigl(\exp(\phi(z))+\exp(-\phi(z))\Bigr),\\ &\Phi_{-1/2}^{1/2}(z^2)=\exp(-\varphi(z^2)) = \frac{z^{-1/2}}2\Bigl(\exp(\phi(z))-\exp(-\phi(z))\Bigr).
\end{align*}
Now we can consider the tensor product $\calL_{1,1}\otimes\calL_{1,1}$. On the top of this representation we have four dimensional representation of $\mathfrak{sl}(2)$ with the basis $\Phi_{1/2}^{1/2}\otimes \Phi_{1/2}^{1/2},\; \Phi_{1/2}^{1/2}\otimes \Phi_{-1/2}^{1/2},\; \Phi_{-1/2}^{1/2}\otimes \Phi_{1/2}^{1/2},\; \Phi_{-1/2}^{1/2}\otimes \Phi_{-1/2}^{1/2}$. This representation can be decomposed as a direct sum of two representations $4=3+1$. The vectors with $h_0=0$, compatible with this decomposition, have the form $\Phi_{1/2}^{1/2}\otimes \Phi_{-1/2}^{1/2}+ \Phi_{-1/2}^{1/2}\otimes \Phi_{1/2}^{1/2}$ and $\Phi_{1/2}^{1/2}\otimes \Phi_{-1/2}^{1/2}- \Phi_{-1/2}^{1/2}\otimes \Phi_{1/2}^{1/2}$. The corresponding local fields in terms of $\phi^{(1)}(z)$ and $\phi^{(2)}(z)$ have the form:
\begin{align*}
& \Phi_{1/2}^{1/2}(z^2)\otimes \Phi_{-1/2}^{1/2}(z^2)+ \Phi_{-1/2}^{1/2}(z^2)\otimes \Phi_{1/2}^{1/2}(z^2)= \frac12\Bigl(\exp(\phi^{(1)}+\phi^{(2)})-\exp(-\phi^{(1)}-\phi^{(2)})\Bigr),\\ &\Phi_{1/2}^{1/2}(z^2)\otimes \Phi_{-1/2}^{1/2}(z^2)- \Phi_{-1/2}^{1/2}(z^2)\otimes \Phi_{1/2}^{1/2}(z^2)= \frac12\Bigl(\exp(\phi^{(1)}-\phi^{(2)})- \exp(-\phi^{(1)})+\phi^{(2)})\Bigr).
\end{align*}
This formulas coincide with \eqref{eq:r2:chir} and \eqref{eq:r2:psir} up to the normalization.

\end{document}